\documentclass[matbio,oneside]{svjour}
\usepackage{graphics}
\usepackage{amssymb}
\usepackage{amsmath}

\begin{document}
\title{Swarming dynamics of a model for biological groups in two dimensions}
\author{Chad M. Topaz \and Andrea L. Bertozzi}
\institute{Dept. of Mathematics, Duke University, Durham, NC 27708 and Dept. of Mathematics, UCLA, Los Angeles, CA 90095; email topaz@math.ucla.edu, bertozzi@math.ucla.edu}
\date{Received: date / Revised version: date}
\maketitle

\begin{abstract} 

We investigate a class of continuum models for the motion of a two-dimensional biological group under the influence of nonlocal social interactions. The dynamics may be uniquely decomposed into incompressible motion and potential motion. When the motion is purely incompressible, the
model possesses solutions which have constant population density and sharp boundaries for all time. Numerical simulations of these ``swarm patches'' reveal rotating mill-like
swarms with circular cores and spiral arms. The sign of the social interaction term determines the direction of the rotation, and the interaction
length scale affects the degree of spiral formation. When the motion is purely potential, the social interaction term has
the meaning of repulsion or attraction depending on its sign. For the repulsive case, the population spreads and the density profile is smoothed. With
increasing interaction length scale, the motion becomes more convective and experiences slower diffusive smoothing. For the attractive case, the population self-organizes
into regions of high and low density. The characteristic length scale of the density pattern is predicted and confirmed by numerical simulations. \end{abstract}

\section{Introduction}
\label{sec:intro}

\subsection{Background}
\label{sec:background}

\begin{figure}
\centerline{\resizebox{\textwidth}{!}
{\includegraphics{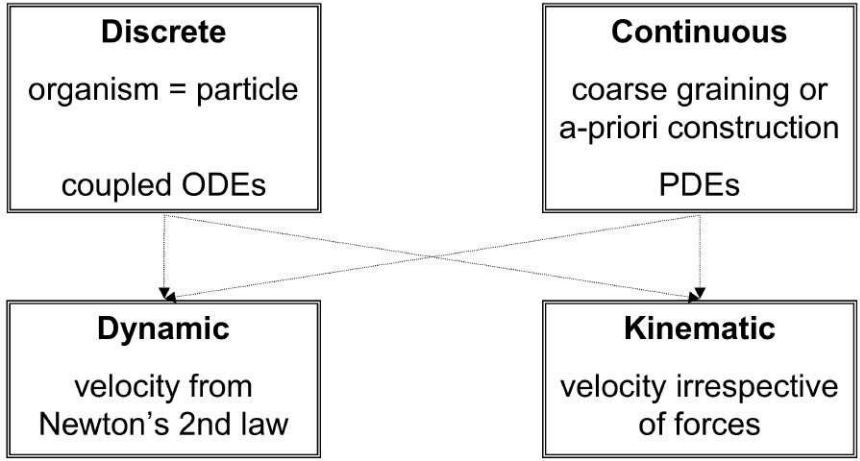}}}\caption{A partial schematic taxonomy of swarming models.}
\label{fig:model-taxonomy}
\end{figure}

Examples of collective motion abound in nature. Swarming, schooling, flocking, and herding have been observed amongst zooplankton, locusts, fish,
birds, wolves, and other organisms; see \cite{ah1990,ph1997,ogk1999} for discussions of such groups. A remarkable aspect of these aggregations is
that individuals move together in a coordinated fashion even though interactions between them via sight, smell, hearing, or other senses are
typically limited to much shorter distances than the size of the group. Over the past few decades, mathematical scientists have begun to tackle the
problem of describing how this coordinated global structure may arise in biological groups. An overview of modeling issues pertaining to swarming is
given in \cite{k2001}, and a partial schematic taxonomy of swarming models, is shown in Figure \ref{fig:model-taxonomy}. We now briefly discuss this
classification of models (noting first that we ignore many other possible categorizations, such as deterministic versus stochastic).

A fruitful approach to modeling swarms has been to treat each individual as a discrete particle. These ``individual-based'' models have been employed
in quite a few biological and mathematical studies, including \cite{vcbcs1995,tt1998,lrc2001,ckjrf2002,ah2003,mkbs2003}. These works begin with simple
rules of motion for each individual, involving some combination of self-propulsion, random movement, and interaction with neighboring organisms. The
models typically take the form of coupled nonlinear difference equations or ordinary differential equations. While numerical simulations of these models have indeed revealed
collective behavior, a principal disadvantage is that for realistic numbers of individuals, analytical results for the collective motion are difficult
or impossible to obtain. It is worth mentioning that some progress has been made in obtaining analytical results for \emph{stationary} groups.
In \cite{mkbs2003}, a discrete model is formulated, and a Lyapunov functional argument is used to successfully predict an equilibrium state of
equally-spaced organisms. However, to our knowledge, analytical (non-statistical) descriptions of non-equilibrium states in discrete swarming models are scarce.

An alternative approach to studying swarms is to focus on continuum equations, which describe relevant quantities as scalar or vector
fields. These models may be constructed a priori, as in \cite{kwg1998,mk1999}, or they may be derived by coarse-graining a particle model, as in
\cite{tt1998,lrc2001}. In general, continuum models provide a convenient setting in which to study large populations since one may apply
machinery from the analysis of partial differential equations (PDEs). In the context of swarming, the focus has generally been on models in which the
population density satisfies a convection-diffusion equation of the form
\begin{equation}
\rho_t + \nabla \cdot (\vec{v}\rho) = \nabla \cdot (D\nabla \rho). \label{eq:c-d}
\end{equation}
Here, $\rho(\vec{x},t)$ is the population density, $\vec{v}(\vec{x},t)$ is the velocity field, and $\vec{x}$ is the (one-, two- or three- dimensional) spatial coordinate. This equation states that the density is conserved while individuals travel with average velocity $\vec{v}$. The motion may involve diffusion, whose strength is measured by $D\equiv D(\vec{x},\vec{v},\rho)$.

Swarming models of the form of (\ref{eq:c-d}) may be classified as either dynamic or kinematic depending on how the velocity field $\vec{v}$ is specified. Dynamic models couple to (\ref{eq:c-d}) an equation for the velocity field, such as
\begin{equation}
\vec{v}_t + \vec{v}\cdot\nabla\vec{v} = \vec{f}(\vec{v}) - \vec{k}(\rho,\vec{v})+\nu\nabla^2\vec{v}+\vec{F}_{ext}. \label{eq:momentum}
\end{equation}
This momentum equation is analogous to Newton's second law. The left-hand side is the material (or convective) derivative, \emph{i.e.} the time derivative in a reference frame moving with the velocity field $\vec{v}$. The right-hand side is simply a sum of forces. The force $\vec{f}$ represents the self-propulsion of individuals, and $\vec{k}$ is a 
nonlocal force due interactions with other members of the population, to be discussed momentarily. The remaining terms on the right side of 
(\ref{eq:momentum}) 
represent a  ``viscosity'' with strength proportional to $\nu$, and an external (environmental) force $\vec{F}_{ext}$. An example of a dynamic model for swarming may 
be found in \cite{lrc2001}. In that work, $\nu = 0, \vec{F}_{ext}=\vec{0}$ and $\vec{f}(\vec{v})=\alpha\vec{v}/|\vec{v}|-\beta\vec{v}$, so that in the 
absence of social interactions, individuals experience a self-propulsion of strength $\alpha$ in their direction of motion and a frictional drag of 
strength $\beta$. 
In contrast, kinematic models, such as the one in \cite{mk1999}, describe the motion of bodies without consideration to the forces acting upon them. That is to say, the velocity does not satisfy a momentum equation, but rather is simply a functional of the population density, \emph{i.e.}
\begin{equation}
\vec{v} = \vec{V}(\rho). \label{eq:kinematic}
\end{equation}
The functional $\vec{V}$ may include effects like those captured in (\ref{eq:momentum}), such as self-propulsion, social interactions, and environmental influence. 

The essence of the swarming phenomenon is the presence of social interactions. For the velocity equation (\ref{eq:momentum}), these
interactions are represented by the term $\vec{k}(\rho,\vec{v})$, and for (\ref{eq:kinematic}), they are contained in the functional $\vec{V}(\rho)$.  
The social interaction terms might describe effects such as attraction or repulsion between individuals sufficiently close to each other, or the
tendency of individuals to orient themselves similarly to their neighbors. Within the context of continuum models, then, the social term takes the
form of an integral operator (most often of convolution form) and the governing equations are actually partial integrodifferential equations (PIDEs).
Continuum models of this form have been studied, for instance, in \cite{lrc2001,mk1999}.

One challenge associated with continuum models has been the difficulty of obtaining biologically realistic swarm solutions, namely, solutions with sharp boundaries, relatively constant internal
population densities, and long lifetimes. For swarms in one spatial dimension, some progress has been made in \cite{mk1999}, which also contains
extensive background and an associated literature review on this issue. We believe that a related issue is the dimensionality of the model. Most
continuum swarm models, such as those in \cite{kwg1998,mk1999,vcfh1999}, have only been investigated in one spatial dimension.  We expect swarming
dynamics in higher dimensional models to be qualitatively different since that case allows for organisms to vary their orientations continuously, as
in the ``mill'' or ``vortex'' states that have been observed in fish schools, ant colonies and other groups \cite{s1971,bccvg1997,pk1999,rnsl1999,obm2003}).

One of our goals in this paper is to highlight a difference in the swarming problem for one and two dimensions. The possibility of rotational motion
in two spatial dimensions allows for cohesive swarms with infinite lifetimes, even when the velocity rule does not include local drift of organisms. Another goal is to demonstrate a natural way of classifying swarming
dynamics in two spatial dimensions, namely by using the Hodge Decomposition Theorem. Our final goal is to make connections between the properties of
the social interaction terms (for instance, their associated length scales and signs) and the large-scale dynamics of the population.

In the following subsection, we mention some results for constant density traveling band solutions of a class of one-dimensional swarm 
models. These are presented for contrast with the two-dimensional case, the study of  which constitutes the bulk of this paper.  In Subsection \ref{sec:intro-part-three}, we state our primary results and outline the remainder of the paper.

\subsection{Elementary results for one-dimensional swarms}
\label{sec:1d-results}

A detailed investigation of a swarming model in one spatial dimension has been carried out in \cite{mk1999}. In that work, the population density is
assumed to obey~(\ref{eq:c-d}). The kinematic velocity rule is \begin{equation} v = a_e \rho + (A_a - A_r \rho)(K*\rho) \label{eq:mkrule} 
\end{equation} Here, the first
term represents a local density-dependent drift. The remaining terms are nonlocal components describing attraction and repulsion, with the asterisk operator having the meaning of convolution. Note that the
repulsive effects are higher order in the the population density than are the attractive ones. The interaction kernel $K$ is odd, piecewise constant, and has compact support. It is given by
 \begin{equation}
K(x) = \begin{cases}
\frac{1}{2d} &\text{if $|x|  \leq r $}\\
0 &\text{otherwise}
\end{cases}
\end{equation}
where $d$ is an interaction length scale parameter which may be freely chosen.

Analysis and numerical simulations in \cite{mk1999} reveal that the model supports swarm solutions with biologically realistic
characteristics, namely a nearly constant internal population density and sharp edges. For density independent diffusion, the cohesive swarm has an exponentially long lifetime before the population is lost through ``tails" in the density profile. For the case of small density-dependent diffusion,
the model has true ``traveling band" solutions which have compact support. In either case, the cohesive motion of the swarm is achieved by an
effective cancellation of the social interactions. The internal density of the swarm is precisely that at which the attractive and repulsive effects
cancel each other, so that the only remaining component of the velocity is local drift. We mention this for contrast with the results to be presented later in this paper, which demonstrate a nonlocal, \emph{i.e.} cooperative, means by which a constant density swarm may move cohesively.

We now mention some simple existence and uniqueness results for one-dimensional swarms with no diffusion. The population density $\rho$ satisfies the
convection equation \begin{equation} \rho_t + \partial_x (v \rho) = 0. \label{eq:convec} \end{equation} We will show how in one-dimension, realistic
velocity rules which are purely nonlocal cannot lead to a constant-speed translation of the population, and thus cohesive swarms cannot be maintained. Again, these results are presented for contrast with the two-dimensional results given later in this paper.

Since we are interested in making statements about
constant density swarms with sharp boundaries, we make the constant density traveling
band (CDTB) \emph{ansatz} \begin{align} \rho(x,t)& =\rho_0 W_L(x-ct)
\label{eq:CDTBansatz1} \\ W_L(x-ct)v(x,t) &= cW_L(x-ct) \label{eq:CDTBansatz2}.  
\end{align} Here, $\rho_0$ is the constant population density, $c$ is the speed of
the traveling band, and $W_L(x)$ is the window function defined without loss of
generality as
 \begin{equation}
W_L(x) = \begin{cases}
1 &\text{if $x \in \Omega_L,\ \Omega_L=[0,L] $}\\
0 &\text{otherwise.}
\end{cases}
\end{equation}
The \emph{ansatz} (\ref{eq:CDTBansatz1}) - (\ref{eq:CDTBansatz2}) automatically satisfies the governing equation (\ref{eq:convec}) in $\Omega_L$, the support of $\rho$. Note that we have not placed any restrictions on the velocity field outside the support of $\rho$ since the velocity in unpopulated areas is irrelevant to the propagation of the swarm. 

We must also consider an equation defining the velocity field. For contrast, we will consider two velocity rules, each of which may be written as a (degenerate) version of the generalized kinematic velocity rule
\begin{equation}
v(x,t) = F(\rho) + G_1(\rho)\left[K_1 * H_1(\rho)\right] +G_2(\rho)\left[K_2 * H_2(\rho)\right]
\label{eq:generalized-velocity}
\end{equation}
This is a generalization of (\ref{eq:mkrule}) from~\cite{mk1999}. $F$ is a functional which captures the local density dependence of the velocity. It represents drift velocity of organisms, irrespective of social forces. $K_1$ and $K_2$ 
are interaction kernels, and thus the $G[K*H]$ terms represent nonlocal effects which arise from repulsion and attraction between organisms. We will assume that $K_1$ and $K_2$ are integrable. In contrast to \cite{mk1999}, we will further assume that these interaction kernels are decreasing in their arguments, so that the influence of the population on a given organism's velocity weakens with distance. We allow the functionals $F,G,H$ to be nonlinear for generality. We assume that $F,G_1,G_2,H_1,H_2$ are smooth, and that $H_1(0)=H_2(0)=0$ so that velocities from social interactions are induced only by nonzero population.

The swarm density $\rho_0$ and the constant band speed $c$ must satisfy a consistency condition via the velocity equation (\ref{eq:generalized-velocity}) in order for the \emph{ansatz} (\ref{eq:CDTBansatz1}) - (\ref{eq:CDTBansatz2}) to be a solution to (\ref{eq:convec}). 

We first consider the case $F=0$, $G_2=0$, so that (\ref{eq:generalized-velocity}) becomes
\begin{equation}
v=G_1(\rho) \left[K_1*H_1(\rho)\right] \label{eq:vrule3}.
\end{equation}
where the argument of $\rho$ is understood to be $z = x-ct$.
Equivalently, we may obtain a rule of this form by choosing $F=0, G_1=G_2, H_1=H_2$, so that
\begin{equation}
v= G_1(\rho) \left[(K_1+ K_2) *H_1(\rho)\right] \label{eq:vrule3alt}
\end{equation}
and both attractive and repulsive effects are represented. Note that for these velocity rules, in the absence of interactions, there is no underlying 
(\emph{i.e.} local) drift velocity.

Combining (\ref{eq:CDTBansatz1}) - (\ref{eq:CDTBansatz2}) and (\ref{eq:vrule3}), we obtain the consistency condition
\begin{equation}
W_Lc=W_LG_1(\rho_0W_L)\left[K_1*H(\rho_0W_L)\right] \label{eq:vrule3mod}.
\end{equation}
which may be rewritten as
\begin{equation}
c = G_1^0 H_1^0 \int\limits_{z-L}^{z}K_1(\zeta)\, d\zeta \quad \text{for $z \in \Omega_L$} \label{eq:case3result}
\end{equation}
where
\begin{equation}
G_1^0 \equiv G_1(\rho_0) \label{eq:G10}
\end{equation}
and 
\begin{equation}
H_1^0 \equiv H_1(\rho_0) \label{eq:H10}
\end{equation}

By differentiating (\ref{eq:case3result}) with respect to $z$ and applying the first fundamental theorem of calculus, we see that
\begin{equation}
K_1(z)=K_1(z-L) \quad \text{for $z \in \Omega_L$} \label{eq:periodicity-condition}.
\end{equation}
Thus, $K_1$ satisfying (\ref{eq:periodicity-condition}) must be $L$-periodic on $[-L,L]$ in order to admit a CDTB solution (the structure of $K_1$ 
outside of $[-L,L]$ is not relevant). We will call the set of such kernels $\Upsilon$. The logical implication goes the reverse way as well, as can be 
seen from writing down a Fourier series for $K_1 \in \Upsilon$, so that  (\ref{eq:case3result}) is satisfied if and only if $K_1 \in \Upsilon$. In 
this 
case, (\ref{eq:case3result}) becomes 
\begin{equation}
c = G_1^0 H_1^0 K_1^a
\label{eq:temp1.7}
\end{equation}
where
\begin{equation}
K_1^a \equiv \int\limits_{0}^{L}K(\zeta)\, d\zeta.
 \label{eq:k1area}
\end{equation}
There may be families of traveling band solutions parameterized by $c$ which bifurcate depending on the structure of the nonlinear functions $G_1$ and 
$H_1$.

However, it is important to realize that the choice $K_1 \in \Upsilon$ is not biologically meaningful, and contradicts our earlier assumption regarding the spatial decay of interaction kernels. As discussed above, biologically realistic kernels are 
expected to satisfy $dK_1/d|z| \leq 0$, so that for a given individual, the effect of other individuals does not increase with distance. 
However, $K_1 \in \Upsilon$ cannot satisfy $dK_1/d|z|<0$, so at best the kernel would be a constant, but even this choice is not expected to be a good biological model, except perhaps for very small $L$.

In contrast to the results just mentioned, we may now consider the case $G_2=0$, so that (\ref{eq:generalized-velocity}) becomes
\begin{equation}
v=F(\rho) + G_1(\rho) \left[K_1*H_1(\rho)\right] \label{eq:vrule4}.
\end{equation}
Equivalently, we may choose $K_1=K_2$, $H_1=H_2$. The velocity rule (\ref{eq:mkrule}) in \cite{mk1999} takes this form. Note that now there is a local drift, a self-induced contribution to the velocity, which is 
captured by $F$. Combining (\ref{eq:CDTBansatz1}) - (\ref{eq:CDTBansatz2}) and (\ref{eq:vrule4}), we obtain the consistency condition
\begin{equation}
W_L c = W_L \left\{ F(\rho_0 W_L)+G_1(\rho_0 W_L) \left[ K_1*H(\rho_0 W_L) \right]  \right\}
\label{eq:vrule4mod} 
\end{equation}
which may be re-written as
\begin{equation}
c = F_1^0 + G_1^0 H_1^0 \int\limits_{z-L}^{z}K_1(\zeta)\, d\zeta \quad \text{for $z \in \Omega_L$}. \label{eq:case4result}
\end{equation}
Here,
\begin{equation}
F_1^0 \equiv F_1(\rho_0) \label{eq:F10}
\end{equation}
and $H_1^0$ and $G_1^0$ are given by (\ref{eq:H10}) and (\ref{eq:G10}).

There are two cases to consider. If $H_1^0 G_1^0 \neq 0$, then (\ref{eq:case4result}) becomes
\begin{equation}
\frac{c-F_1^0}{G_1^0 H_1^0} = \int\limits_{z-L}^{z}K_1(\zeta)\, d\zeta \quad \text{for $z \in \Omega_L$}.
\label{eq:case4cond1}
\end{equation}
This is similar to the previous case. The condition (\ref{eq:case4cond1}) may be met only for $K_1 \in \Upsilon$, in which case existence and 
uniqueness of solutions depends on the structure of $(c-F_1^0)/(G_1^0 H_1^0)$. For $K_1 \notin \Upsilon$, CDTB solutions are not possible.

On the other hand, if $H_1^0 G_1^0 = 0$, then CDTB are possible for \emph{any} choice of kernel $K_1$. In this case, the number of possible CDTB 
solutions depends on the number of roots of  $H_1^0 G_1^0$ for positive $\rho$, of which there are expected to be a finite number. Looking at the 
problem from the forward (rather than inverse) perspective, for biologically realistic choices of $K_1$, the velocity rule (\ref{eq:vrule4}) will lead 
to a finite number of CDTB solutions. The densities correspond to the solutions $\rho_0^* > 0$ of $H_1^0 G_1^0=0$, and the wave speeds are given by 
$F_1(\rho_0^*)$. Thus, the combination of local and nonlocal velocity terms selects particular densities and band speeds, rather than allowing entire 
families of solutions, as in the purely nonlocal case. The allowed CDTB densities are those at which the nonlocal interactions disappear. Further, 
since we imagine the total population to be fixed in number, this velocity rule actually dictates preferred swarm sizes~$L$. These conclusions are 
similar to those reached in \cite{mk1999} for the particular choice of $F_1,G_1,H_1,K_1$ given by~(\ref{eq:mkrule}).

\subsection{Outline and summary}
\label{sec:intro-part-three}

The remainder of this paper is devoted to an examination of a nonlocal kinematic swarming model in two spatial dimensions. In Section
\ref{sec:2d} we formulate an abstract model of animal motion based on simple assumptions. We also discuss how the Hodge Decomposition Theorem
provides a useful way of understanding the two-dimensional motion of the group, namely by decomposing it into incompressible motion and
potential motion.

In Section \ref{sec:incomp} we focus on the case of incompressible motion. Since we wish to study the motion of a biologically-realistic swarm, we
assume the initial condition to be a finite group with constant internal population density and sharp edges. We show that such a swarm
retains these characteristics for all time. Numerical simulations demonstrate that the dynamics of the swarm are rotational, and that the asymptotic
states are vortex-like structures with circular cores and a potentially complex arrangement of spiral arms. The sign of the social interaction term
determines the direction of rotation of the swarm, and the characteristic length scale of the interactions determines the degree of spiral formation.
The spiral states we observed are qualitatively similar to the mill states observed in \cite{s1971,bccvg1997,pk1999,rnsl1999,obm2003}).

In Section \ref{sec:potential} we consider the complementary case of potential motion. The sign of the social interaction term determines
whether the interaction represents nonlocal repulsion or attraction. These effects lead, respectively, to dispersion or aggregation of the
population. For the dispersive case, shorter interaction length scales result in smoothing of the population density profile, while larger
interaction length scales lead to motion which is more convective. For the case of aggregation, a simple linear stability analysis enables us to
identify a most unstable wavelength, and thus make a prediction about the characteristic length scale of the clumped population distribution that will form. We demonstrate
these results by means of numerical simulations.

Finally, we conclude in Section \ref{sec:conclusions} with a brief summary and a discussion of directions for future investigation.

\section{A kinematic two-dimensional swarm model}
\label{sec:2d}

For the remainder of this paper, we study the dynamics of a two-dimensional swarming model. In choosing our model, we make the following assumptions:
\begin{enumerate}
\item{The population density is conserved; birth, death, immigration, and emigration of organisms are negligible on the time scale of the 
swarming dynamics.}
\item{The motion of organisms is due solely to social interactions, and thus velocities depend nonlocally on the population density (no drift).}
\item{Interactions between organisms are pairwise.}
\item{The social interactions are a linear functional of the population density.}
\item{The social interactions depend only on the distance between organisms, and become weaker with increasing distance.}
\end{enumerate}
Implicit in the second assumption is the supposition that random movement (\emph{e.g.} due to fluctuations in the organisms' medium or noise in their 
ability to move) is negligible. The third and fourth, and fifth assumptions are made for tractability of the model. The third assumption is made so that interaction effects on a given organism will be summable, and this will lead to a convolution in our model, similar to the model in \cite{mk1999}.

In the spirit of the work in \cite{mk1999}, we construct an abstract model, and thus we do not incorporate many biological specifics. Our
model might be interpreted as a one for ``flat" (two-dimensional) groups in the absence of disturbances such as predators or food sources. Even with the simple
assumptions we have made, the dynamics are complex. As discussed in Section \ref{sec:conclusions}, relaxing some of our assumptions to obtain a
more biologically realistic model will be an element of our future work.

Under the assumptions described above, the model takes the form
\begin{gather}
\rho_t + \nabla \cdot (\vec{v}\rho) = 0 \label{eq:2dmodel1} \\
\vec{v} = \int\limits_{\mathbb{R}^2} \vec{K}(|\vec{x}-\vec{y}|)\rho(\vec{y})\, d\vec{y} \equiv \vec{K} * \rho. \label{eq:2dmodel2}
\end{gather}
Here,  $\vec{x}=(x,y)$ is the two-dimensional spatial coordinate. Note that (\ref{eq:2dmodel1}) is simply (\ref{eq:c-d}) with $D=0$, and (\ref{eq:2dmodel2}) is a two-dimensional analog of a degenerate case of the velocity rule (\ref{eq:generalized-velocity}). $\vec{K}$ is our two-dimensional social interaction kernel, which is spatially-decaying and isotropic.

Since our model includes no drift term, velocities decay in the far field and we may apply the Hodge Decomposition Theorem (see, for 
instance,~\cite{mb2002}). This theorem states that a vector field in the plane may be uniquely decomposed into a divergence-free component and a gradient
component. That is to say, the velocity may be written as
\begin{equation}
\vec{v} = \psi + \nabla \Phi,\quad \nabla \cdot \psi = 0.
\end{equation}
For smooth vector fields decaying at infinity, the divergence free part has a scalar stream function $\Psi$ satisfying $\nabla^\perp\Psi = 0$.
Thus, we can write 
\begin{equation}
\vec{v} = \nabla^\perp \Psi + \nabla \Phi.
\end{equation}

Using an analogy to fluid flow, we may think of $\Psi$ as a stream function for the incompressible part of the flow and $\Phi$ as a pressure due
to interactions. For functions with integrable gradients, convolution commutes with derivatives, i.e. $(\nabla P) * \rho = \nabla (P*\rho)$, so
that for the model (\ref{eq:2dmodel1}) - (\ref{eq:2dmodel2})  we can directly apply the Hodge decomposition to the interaction kernel $\vec{K}$:
\begin{equation} \vec{K} = \nabla^\perp N + \nabla P \end{equation} where $P$ models the interaction pressure (motion towards and away from
concentrations of density) and N models additional motion which, as we will see, allows for rotation and a cohesive swarm.

To better understand the model (\ref{eq:2dmodel1}) - (\ref{eq:2dmodel2}), we separate the dynamics into the two cases which we have just
discussed, namely incompressible motion and potential motion.  In the following two sections we study each case in turn, and demonstrate how
the macroscopic motion of the population is affected by the interaction kernel $\vec{K}$.

\section{Incompressible motion}
\label{sec:incomp}

In this case, \begin{equation} \vec{K}=\nabla^{\perp}N \label{eq:incompkernel} \end{equation} so that $\nabla \cdot \vec{v} = 0$. Note that this rule 
has no meaning in one spatial dimension, since there is no notion of perpendicular movement. In two dimensions, we will see that this type of 
interaction allows for cohesive movement of the swarm \emph{without} a local drift term.

The governing equations (\ref{eq:2dmodel1}) - (\ref{eq:2dmodel2}) may 
be written compactly as
\begin{equation}
\rho_t + \nabla \cdot [\rho(\nabla^{\perp}N * \rho)] = 0. \label{eq:2dincomp}
\end{equation}
We take the scalar interaction function $N$ to be a Gaussian of width $d$, \emph{i.e.}
\begin{equation}
G_d(|\vec{x}|) \equiv \frac{1}{d^2} e^{-|\vec{x}|^2/d^2} \label{eq:Gaussian}
\end{equation}
One might include an additional constant prefactor, but this would represent a velocity scale and may be removed by rescaling the time variable in the equations. The length scale $d$ could also be removed by rescaling, in which case the only parameter remaining in the problem would be the initial condition. We choose to retain the length scale parameter $d$ since it has a clear biological interpretation, and since it is more convenient to vary than the length scale of the initial condition.

A Gaussian interaction was also considered for a linear stability analysis in \cite{mk1999}. Other works have used power functions or decaying exponentials \cite{lrc2001,mkbs2003}.  Our interaction function has somewhat different meaning than the ones used in these previous works since it will be applied in two dimensions and thus has a rotationally-symmetric structure.
We choose Gaussian interaction functions since they are biologically realistic (in terms of being spatially decaying) and because they have convenient mathematical properties such as bounded norms and infinite differentiability. We do not intend for this choice to be taken too literally as approximating pairwise interactions. Certainly other choices of kernels would be equally plausible. Nonetheless, many of our qualitative results will hold true for other classes of smooth, spatially decaying interaction functions with normalized integral.

We will begin by making some general statements about the effect of varying the interaction length scale $d$. For very small values of $d$, the interaction
function $N$ resembles a $\delta$-function of strength $\pi$. For the limiting case $d \rightarrow 0$,~(\ref{eq:2dincomp}) may be written as \begin{equation}
\rho_t + \pi \nabla \cdot [\rho\nabla^{\perp} \rho)] = 0 \label{eq:2dincomp2} \end{equation} since $\nabla^{\perp}$ commutes with
convolution and the $\delta$-function acts as the identity under convolution. A little algebra reveals that (\ref{eq:2dincomp2}) is actually
$\rho_t=0$, and thus the swarm will be stationary. This makes intuitive sense. In the case that motion is perpendicular to population gradients in
a completely local sense, the population density profile cannot change, by construction. Of course, in a Lagrangian frame (tracking the coordinates of
an individual organisms) motion is possible, as long as it is perpendicular to the gradient.

On the other hand, for very large values of $d$, $N$ is nearly a constant, namely zero. In the formal limit $d \rightarrow \infty$, $\nabla^{\perp}N = 0$, and once again (\ref{eq:2dincomp}) becomes
$\rho_t=0$. This result also makes intuitive sense. In this case, organisms can sense population gradients infinitely far away, but these gradients have no influence on velocity since social interactions are infinitely weak, and thus the organisms are stationary.

For simplicity, and for analogy with the results mentioned in Section~\ref{sec:1d-results}, we now focus on constant density solutions of compact
support. That is to say, we assume that the initial condition is a \emph{swarm patch} with finite area and constant population density $\rho_0$. By
making such a choice, we are not modeling the initial formation of a constant-density swarm. Rather, this model should be interpreted as a macroscopic
description of a swarm in which attractive and repulsive forces have already come into balance. We will study the subsequent movement of such a swarm.

We use Green's formula to rewrite (\ref{eq:2dmodel1}) - (\ref{eq:2dmodel2}) as an integral over the boundary:
\begin{equation}
\vec{v}(\vec{x}) = \rho_0 \int\limits_{\partial\Omega}N(|\vec{x}-\vec{y}|)\vec{t}(\vec{y})ds(\vec{\vec{y}}).
\label{eq:condyn}
\end{equation}
where $\Omega$ is the support of $\rho$, and the boundary $\partial\Omega$ is parameterized in a clockwise orientation. Here $s$ is the arc length and
$\vec{t}$ is the unit tangent vector to the boundary. Following the strategy from fluid dynamics, we adopt a Lagrangian framework and
track points on the boundary of the swarm patch. That is to say, we write down a 
Langrangian formulation of~(\ref{eq:condyn}) which will be useful for numerical simulations. Taking $\alpha$ to parameterize
the boundary of the swarm, we have the equation for $\vec{z}(\alpha,t)$, the patch boundary:
\begin{equation}
\frac{d\vec{z}(\alpha,t)}{dt} = \int_0^{2\pi} N(\vec{z}(\alpha,t) - \vec{z}(\alpha',t)) \vec{z}_{\alpha}(\alpha',t) d\alpha'. \label{eq:cde}
\end{equation}
where the subscript $\alpha$ indicates a derivative along the boundary.

This equation describes a self-deforming curve. From a computational standpoint, this formulation is convenient because the dimension of the problem has been reduced by one. More importantly, we see that since the boundary is a self-deforming curve, \emph{the swarm patch retains constant internal density and compact support for all time}. The philosophy here is similar to the contour-dynamics formulation of
the two dimensional Euler equations \cite{zhr1979} which describe how a fluid region of constant vorticity, or \emph{vortex patch}, evolves in time. The
difference is that for the swarm patch case, the interaction function $N$ is expected to be spatially decaying in order to be biologically
meaningful (\emph{cf.} our modeling assumptions at the start of Section \ref{sec:2d}, and our choice in equation \ref{eq:Gaussian}) while for the
vortex patch problem, $N = \frac{1}{2\pi} \log|\vec{x}|$.

Before presenting numerical results of this model, we remark on the smoothness of the swarm boundary. For the case of fluid dynamical vortex
patches mentioned above, solutions of (\ref{eq:cde}) with smooth initial data are known to stay smooth for all time \cite{bc1993,c1994}.  This
is the case for our present swarm patch problem as well. See the appendix for the sketch of a proof.

Equation (\ref{eq:cde}) may be solved numerically to find the evolution of the swarm patch boundary. We now briefly describe our simple numerical algorithm. An initial swarm patch shape is selected, and the boundary is discretized into $n$ nodes. Depending on our initial condition, we take the initial number of Lagrangian nodes to be between $n=40$ and $n=60$. The shape of the patch is evolved by using the discretized version of (\ref{eq:cde}). The position of each node may be updated by computing its velocity and then using a time-stepping rule. We perform the spatial integral in (\ref{eq:cde}) using Simpson's rule, which is an $\mathcal{O}(n^2)$ operation. As a start-up procedure, we take three time steps using a fourth order Runge-Kutta method. However, since the Runge-Kutta method involves many evaluations of the right-hand side of (\ref{eq:cde}), it is computationally expensive. Thus, we use a fourth-order multi-step Adams-Bashforth rule for the remainder of the time steps. We take a time step of size $\Delta t = 0.02$. Checks are performed with smaller time steps and varying initial discretizations of the swarm patch boundary to verify that our solutions are sufficiently well-converged.

Despite the fact that the boundary stays smooth, numerical simulations reveal that it develops complex structure (as we show below) which is
also a feature of vortex patches \cite{d1989}. As the swarm patch evolves, it may be necessary to re-discretize the boundary in order to have an
accurate solution (\emph{i.e.} to have a fine enough mesh to capture new spatial complexity).  We do so at every time step. Nodes which are
adjacent with respect to the Lagrangian parameter $\alpha$ are checked for spacing. If the Euclidean distance becomes too large, a node is
inserted between them using linear interpolation. Similarly, if nodes become too close together, they are replaced with a node whose position is
the spatial average of the original ones. While this latter step discards detail below a certain length scale, we perform it nonetheless so that
the total number of nodes does not grow so quickly as to make the computation prohibitively slow. Finally, we periodically perform a check to
verify that the swarm-patch boundary is not self-intersecting (self-intersection of the boundary would break the uniqueness of particle paths which the problem must obey). If the boundary is found to self-intersect, the simulation is aborted, and must be repeated with a finer threshold of spatial detail.

\begin{figure}
\centerline{\resizebox{\textwidth}{!}
{\includegraphics{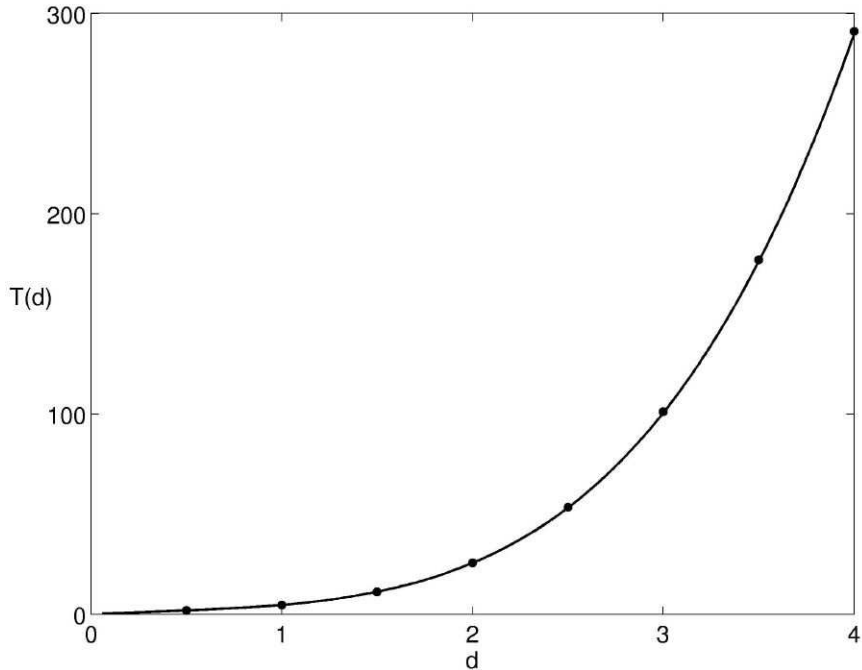}}}\caption{Period of rotation $T$ for the boundary of a circular swarm patch under the model (\ref{eq:2dincomp}) with the  scalar interaction function (\ref{eq:Gaussian}). Here, $T$ is shown as a function of $d$, the interaction length scale in (\ref{eq:Gaussian}). The line corresponds to the exact expression $T=2\pi R/|\vec{v}(R)|$ given by (\ref{eq:circleperiod}). The dots correspond to a numerical simulation of the contour dynamics equation (\ref{eq:cde}). For this example, the radius of the patch is $R=1$ and the constant population density is $\rho_0=1$.}
\label{fig:circleperiod}
\end{figure}

We note that by symmetry arguments, a rotating disk is an exact solution to (\ref{eq:2dincomp}) (though the rotation is not solid-body
rotation). This is true for fluid vortex patches as well; see \cite{mb2002} for a discussion. We assume that $N$ is the Gaussian interaction
function given by (\ref{eq:Gaussian}) and calculate the resulting velocity of points on the swarm boundary, a circle of radius $R$. After some
algebra, we find the velocity $|\vec{v}(R)|$, from which we may compute the period of rotation \begin{equation} T(D)=\frac{2\pi R}{|\vec{v}(R)|} =
\frac{d^2}{\rho_0}e^{2R^2/d^2} \left\{I_1(2R^2/d^2) \right\}^{-1} \label{eq:circleperiod} \end{equation} where $I_1$ is the modified Bessel
function of the first kind of order one. Figure~\ref{fig:circleperiod} shows the period of rotation of the boundary, $T(d)$, for a swarm patch
of radius $R=1$ and population density $\rho_0=1$. The exact expression~(\ref{eq:circleperiod}) is plotted as a line, while data from a
numerical simulation, obtained by tracking one of the Lagrangian nodes, is plotted as dots. These results not only serve as a check on our
algorithm, but demonstrate our previous conclusions about the the behavior of (\ref{eq:2dincomp}) in the limits $d \rightarrow 0$ and $d \rightarrow \infty$ for the particular case of a circular patch.

This example also touches on a connection between the interaction function $N$ and the direction of rotation. Since the boundary of the patch
has a direction associated with its parameterization, the rotation is clockwise or counterclockwise according to whether the interaction
function $N$ is chosen to have, respectively, a positive or negative sign. This limitation results from our choice of kinematic velocity rule. In the
case of a dynamic velocity rule, inertial effects would give any initial swarm patch a natural direction of rotation. That is to say, for a dynamic rule, the swarm will have the freedom to nucleate a rotational state, for instance, as seen in simulations of the model in \cite{lrc2001}. In this section, for our
kinematic velocity rule, we choose $N$ to have a positive sign, and thus swarm patches will always rotate in a clockwise manner. 

We now turn to a discussion of the behavior of the model for other (noncircular) initial conditions, and for intermediate values of $d$ when
some nontrivial evolution occurs. We find that for the present case of incompressible velocity, the dynamics are characterized by an overall
rotational motion. The solutions at sufficiently long times are vortex-like, as we now illustrate with several examples.

\begin{figure}
\centerline{\resizebox{\textwidth}{!}{\includegraphics{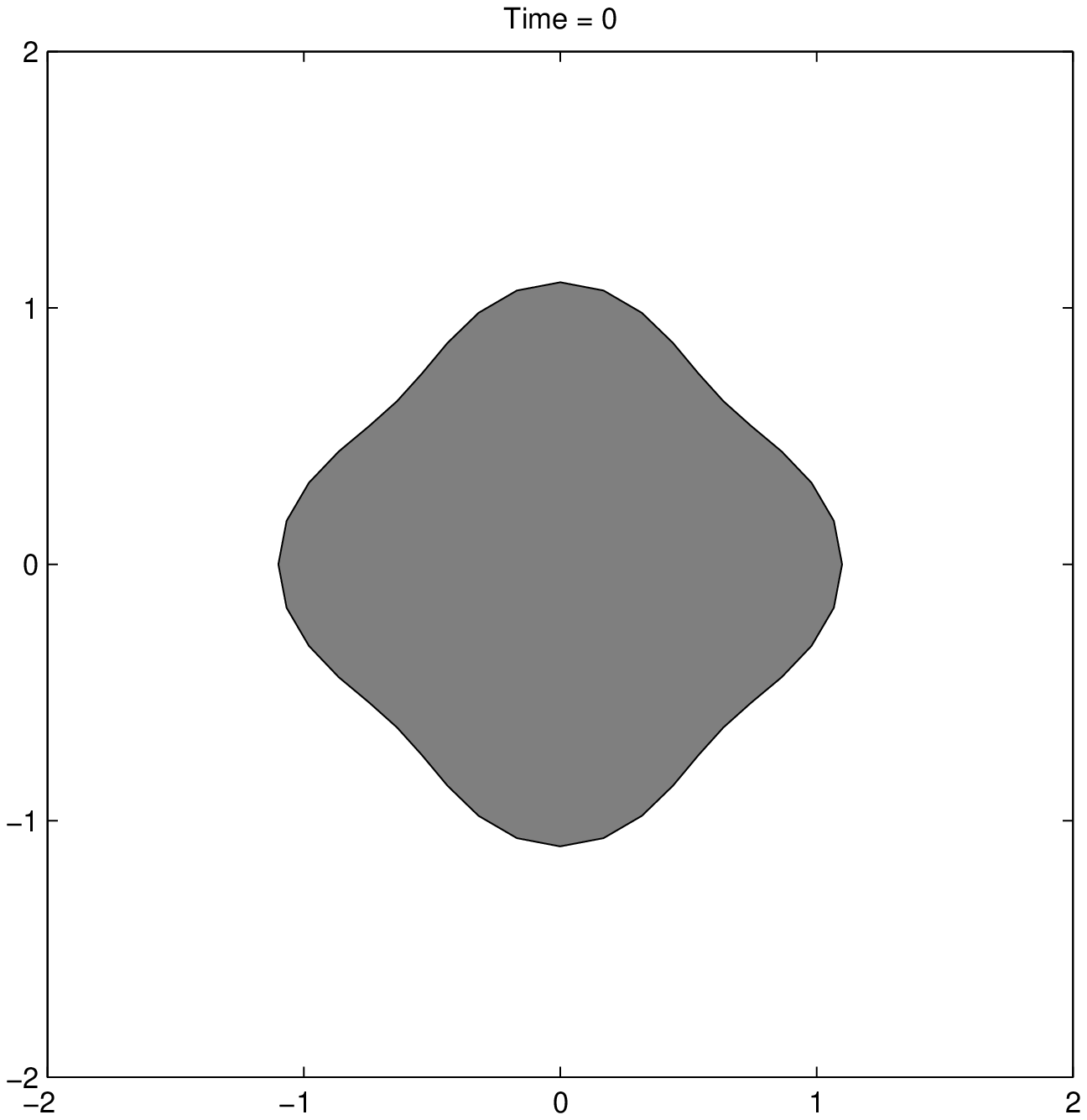} \hfill \includegraphics{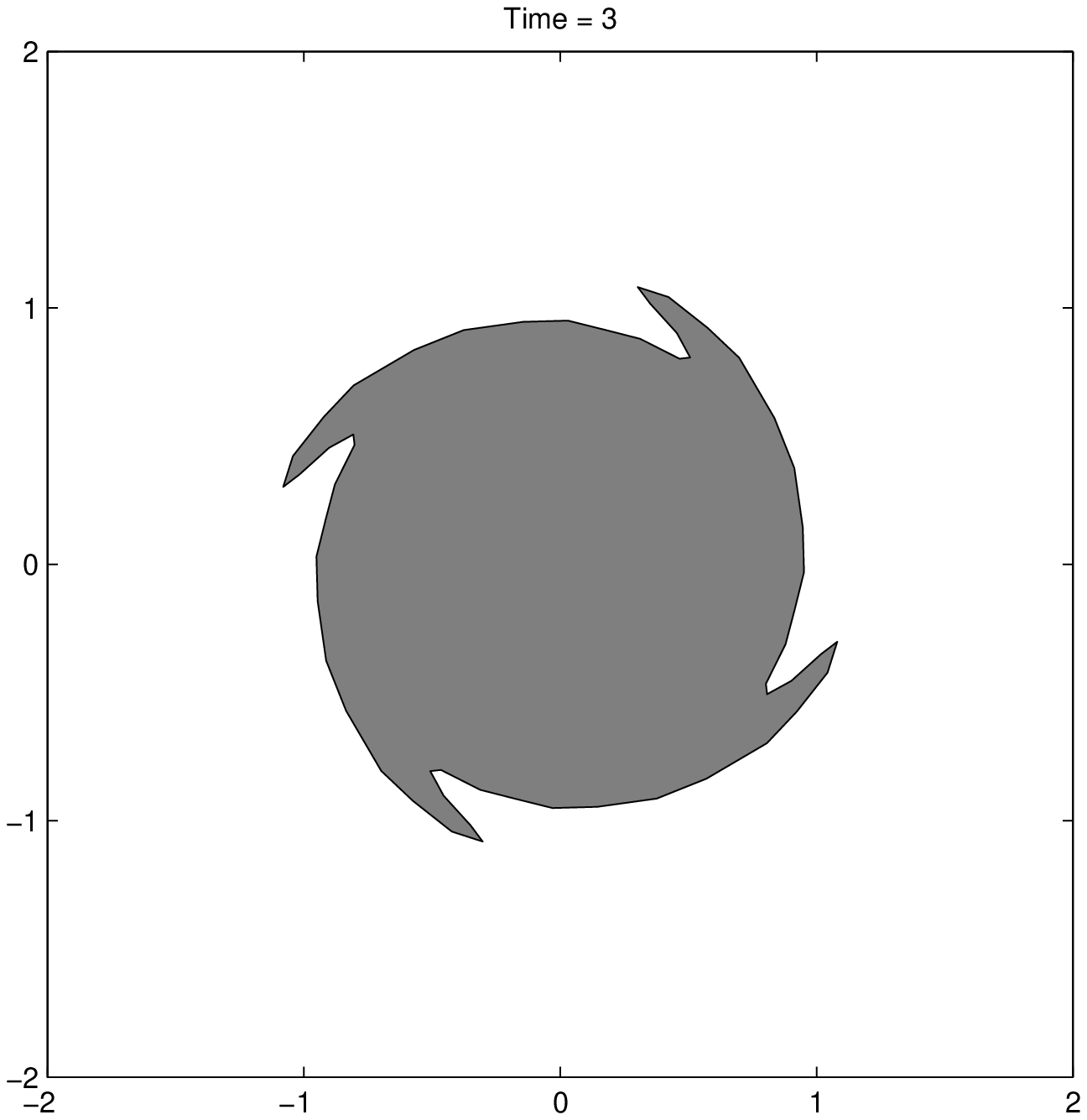}}}
\centerline{\resizebox{\textwidth}{!}{\includegraphics{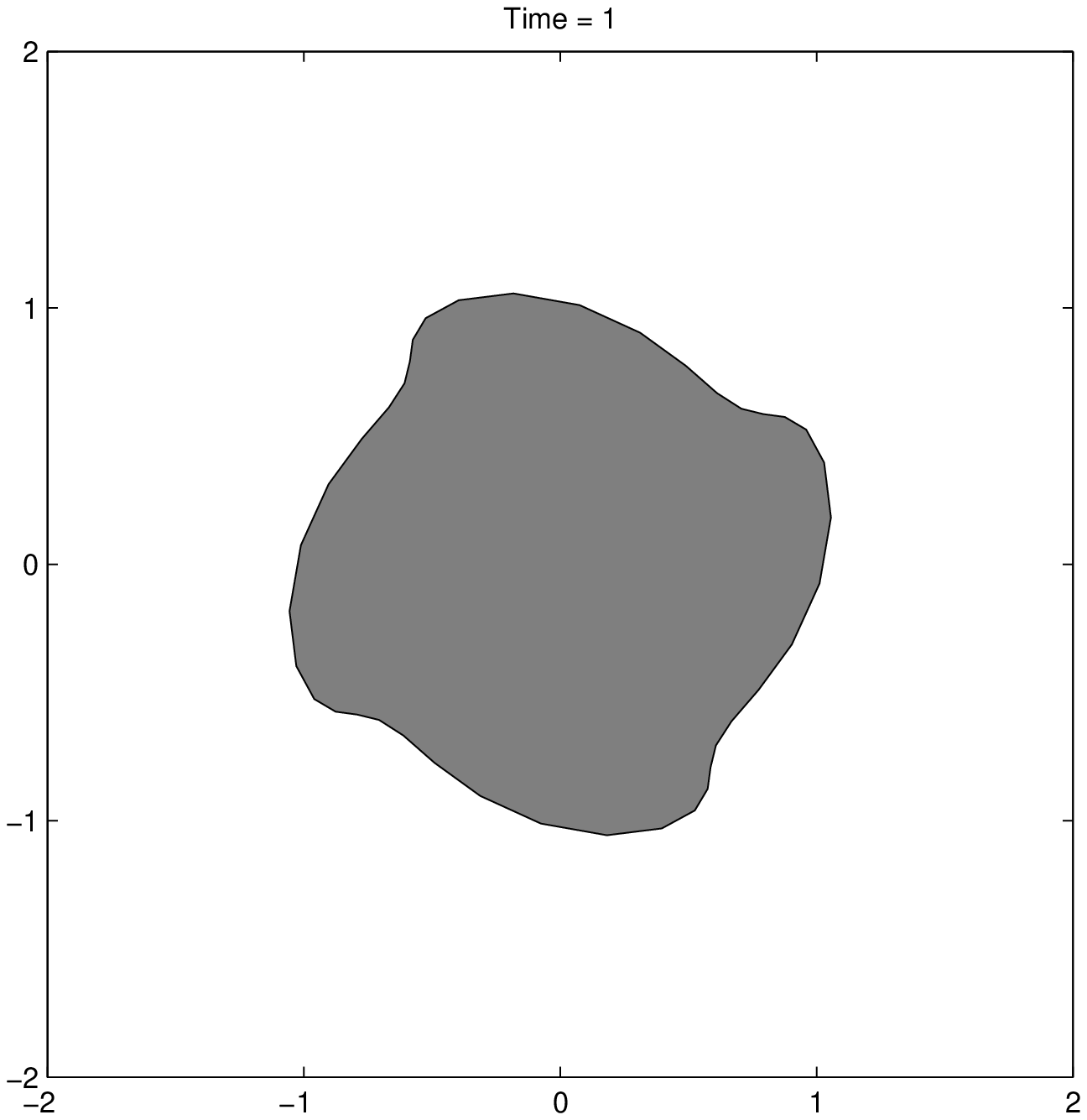} \hfill \includegraphics{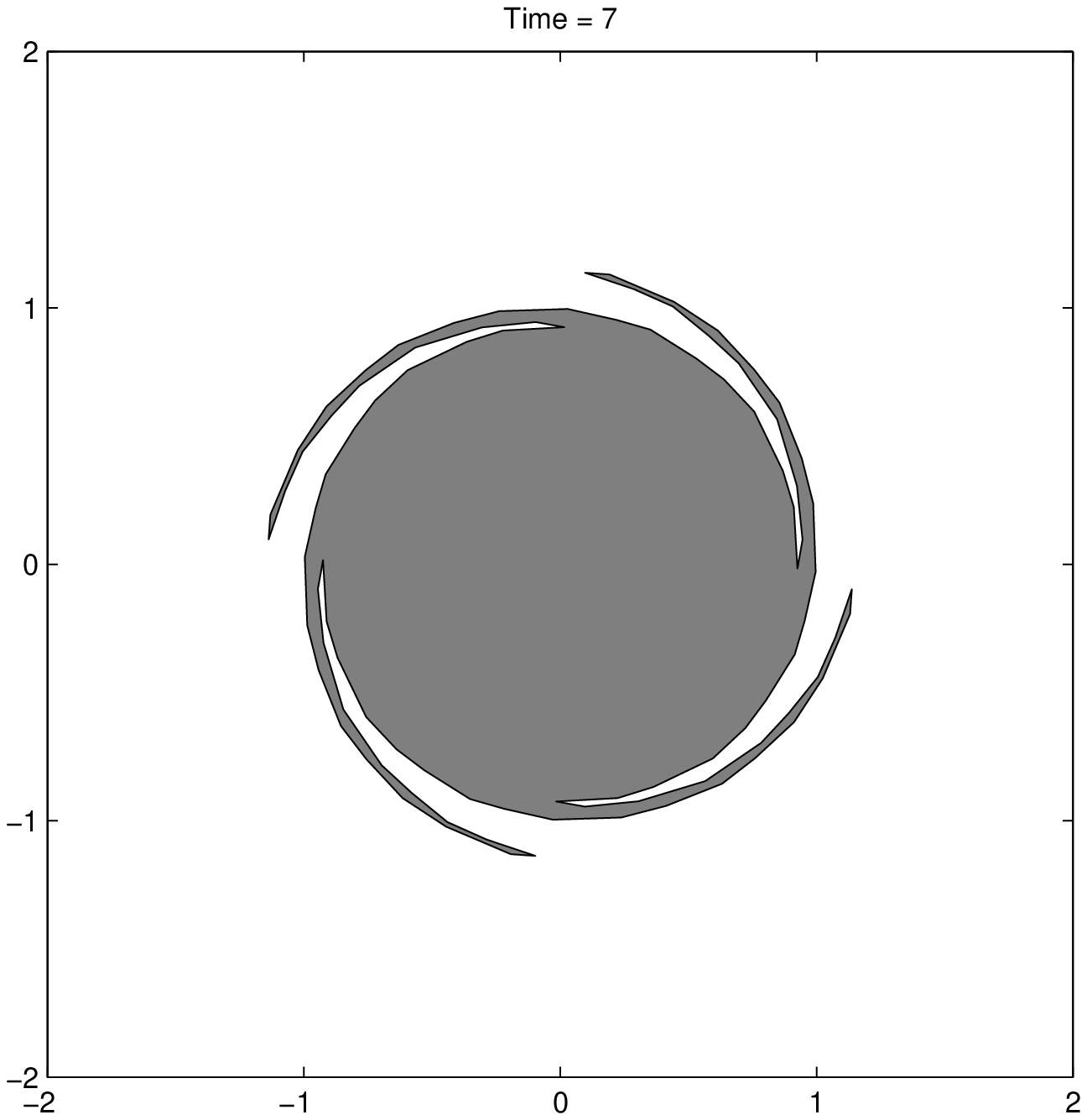}}}
\centerline{\resizebox{\textwidth}{!}{\includegraphics{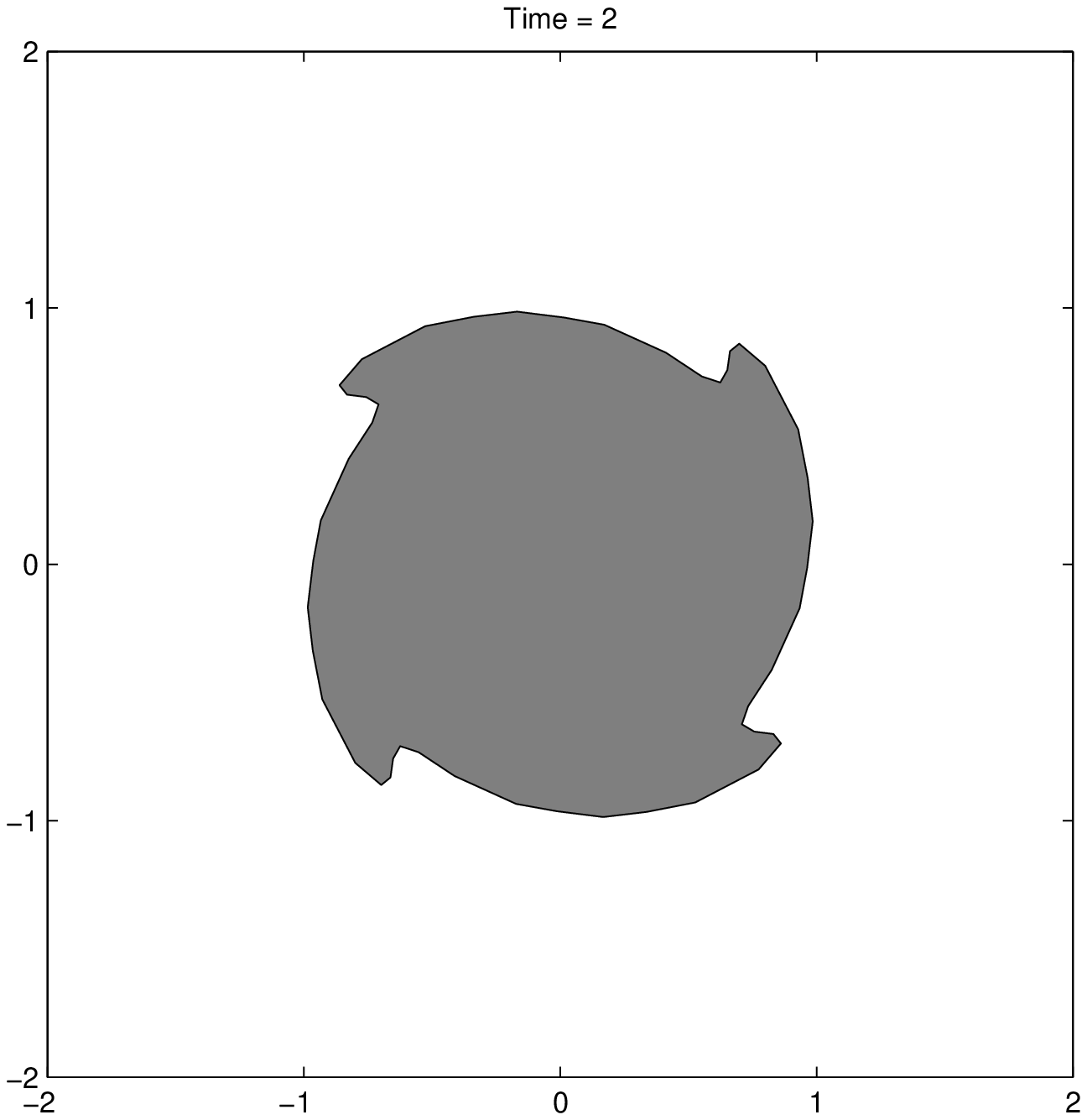} \hfill \includegraphics{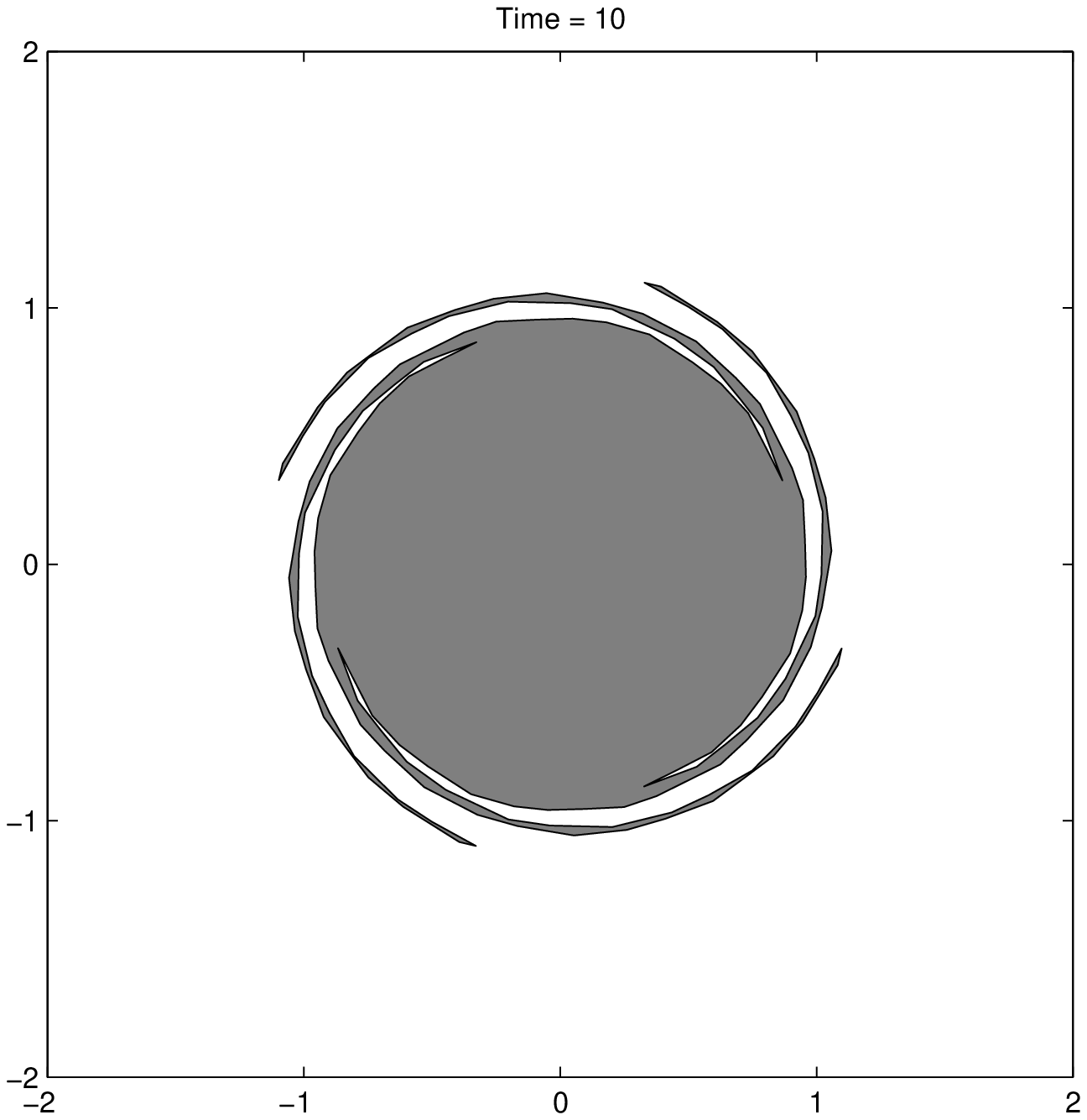}}}
\caption{Evolution of a swarm patch under the model (\ref{eq:2dincomp}). The boundary of the initial shape is a polar curve with radius given by $r(\theta) = 1+(1/10)(\cos 4\theta)$. The constant population density is $\rho_0 = 1$.The  scalar interaction function $N$ is the Gaussian given by (\ref{eq:Gaussian}) with interaction length scale $d=1$. The constant density swarm patch rotates clockwise and develops spiral arms.}
\label{fig:squareseries}
\end{figure}

Figure \ref{fig:squareseries} shows the evolution of a swarm patch using the interaction function (\ref{eq:Gaussian}) with $d=1$. The initial boundary of the population is the polar curve $r(\theta) = 1+(1/10)(\cos 4\theta)$. The square-like initial swarm patch experiences clockwise rotation. At time $t = 1$, the beginnings of spiral arms are visible at the corners of the patch, where the initial curvature was greatest. By time $t=3$, the spirals have grown longer and the core of the patch is becoming circular. This trend continues through the end of the simulation at $t=10$, at which point the spiral arms have grown even longer and the core is nearly a perfect circle.

\begin{figure}
\centerline{\resizebox{\textwidth}{!}{\includegraphics{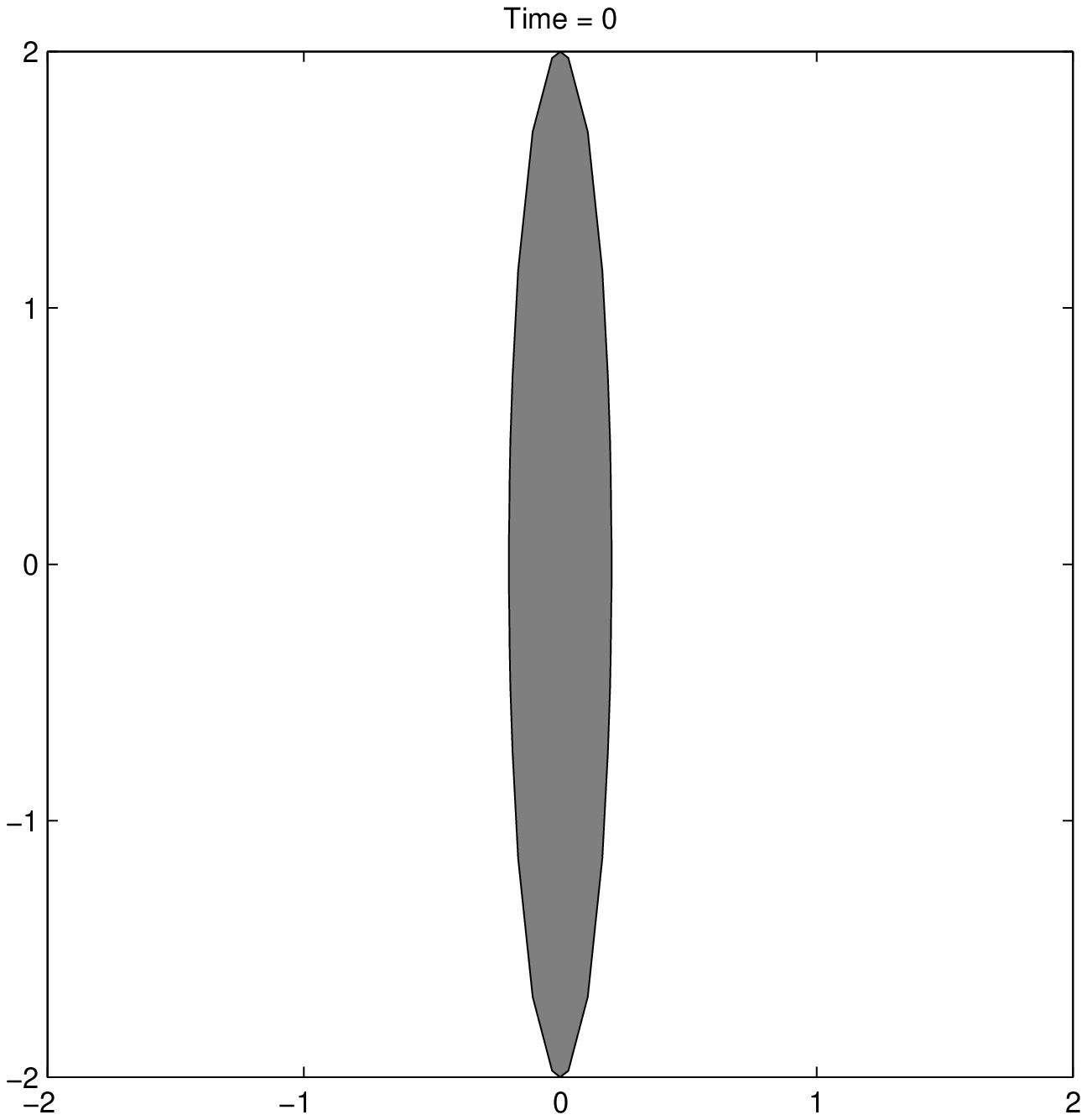} \hfill \includegraphics{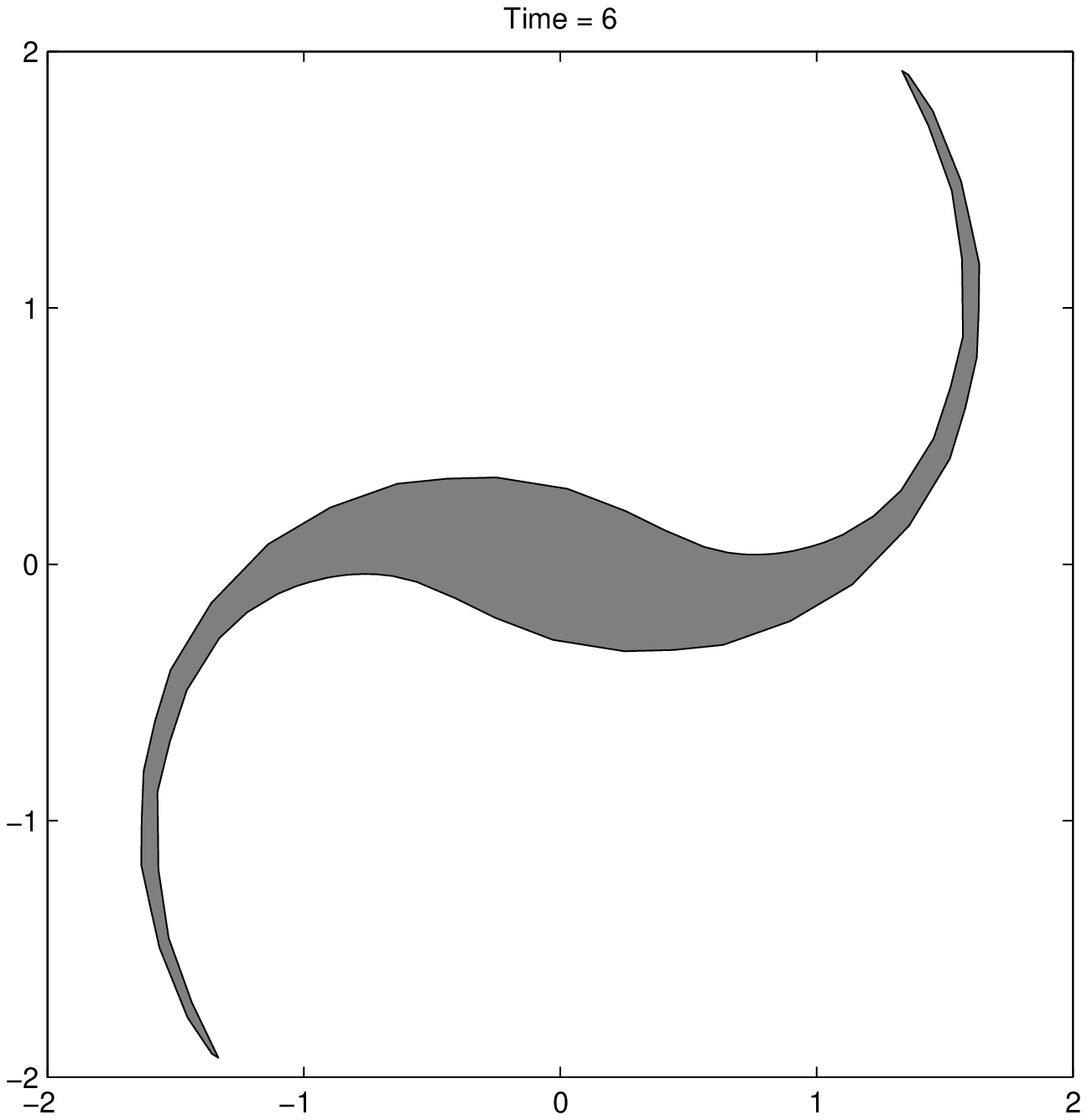}}}
\centerline{\resizebox{\textwidth}{!}{\includegraphics{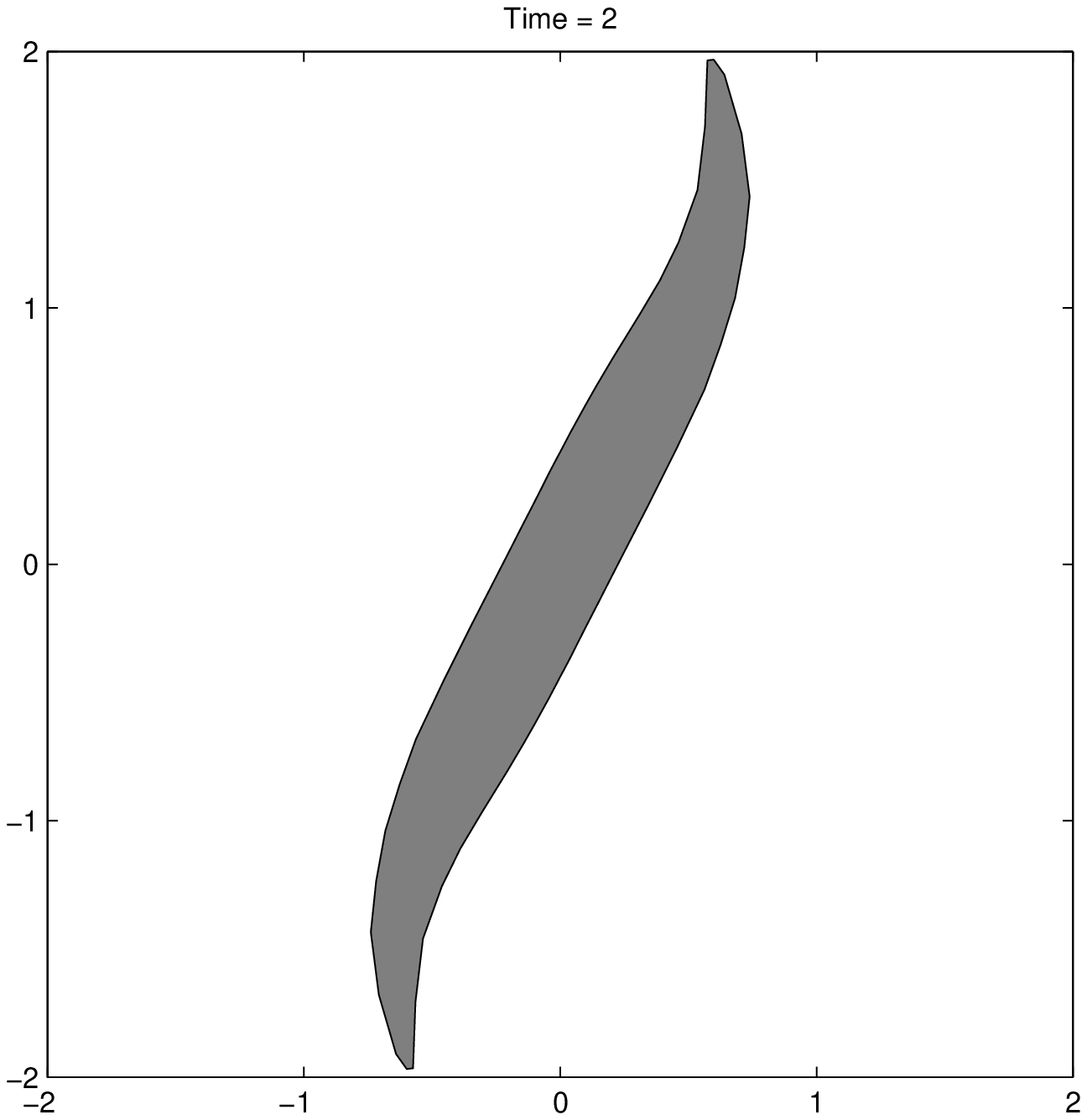} \hfill \includegraphics{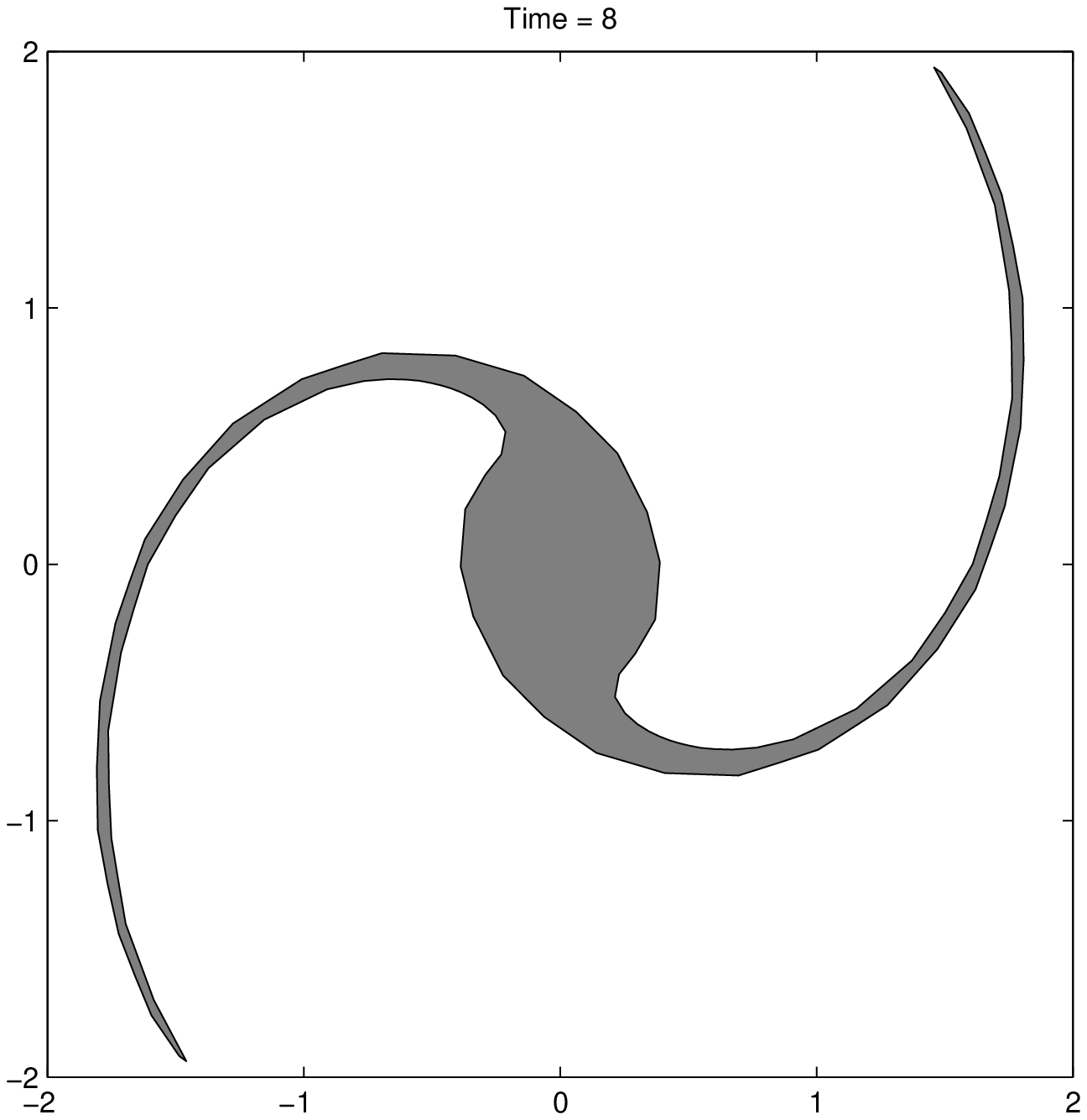}}}
\centerline{\resizebox{\textwidth}{!}{\includegraphics{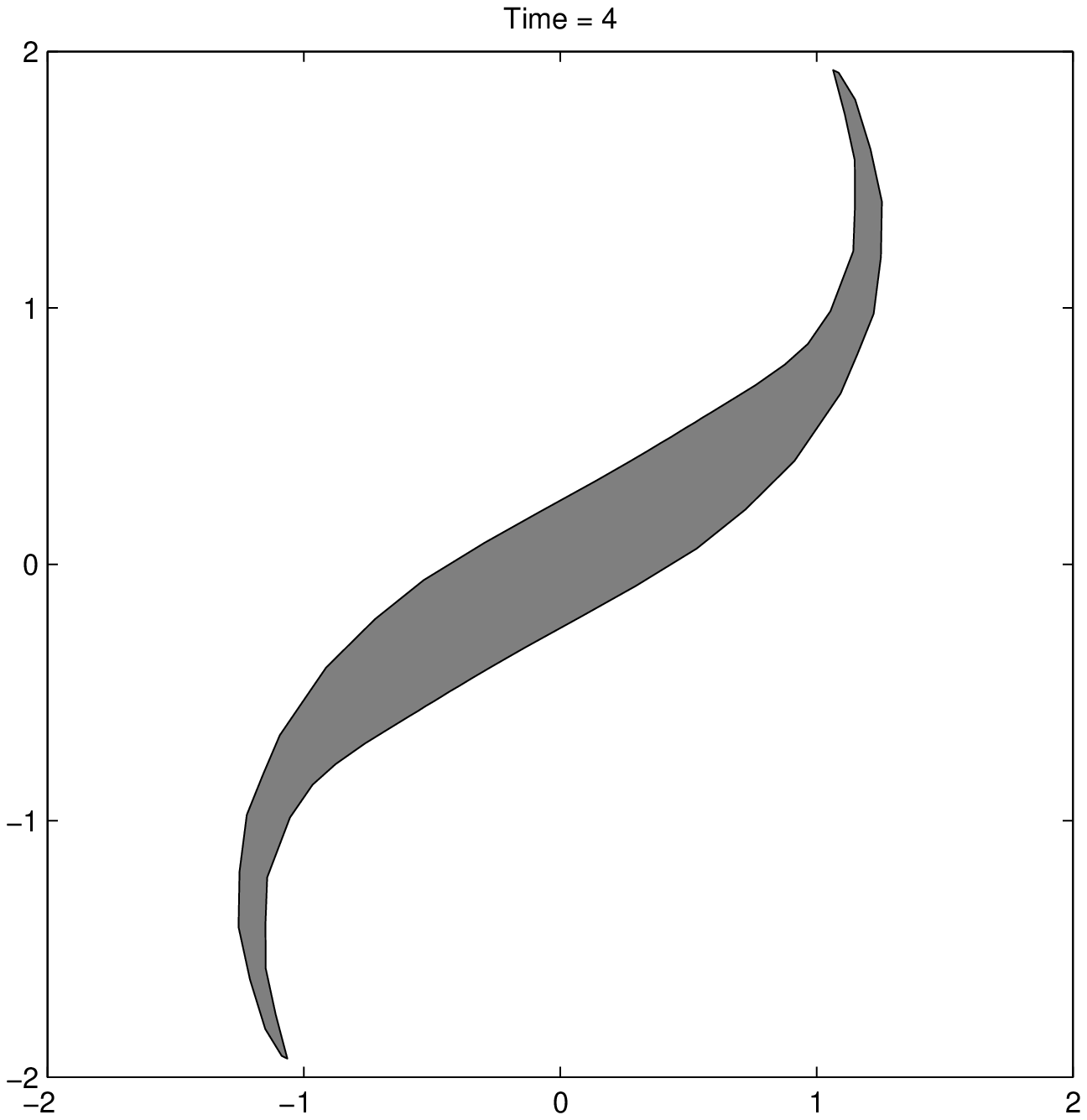} \hfill \includegraphics{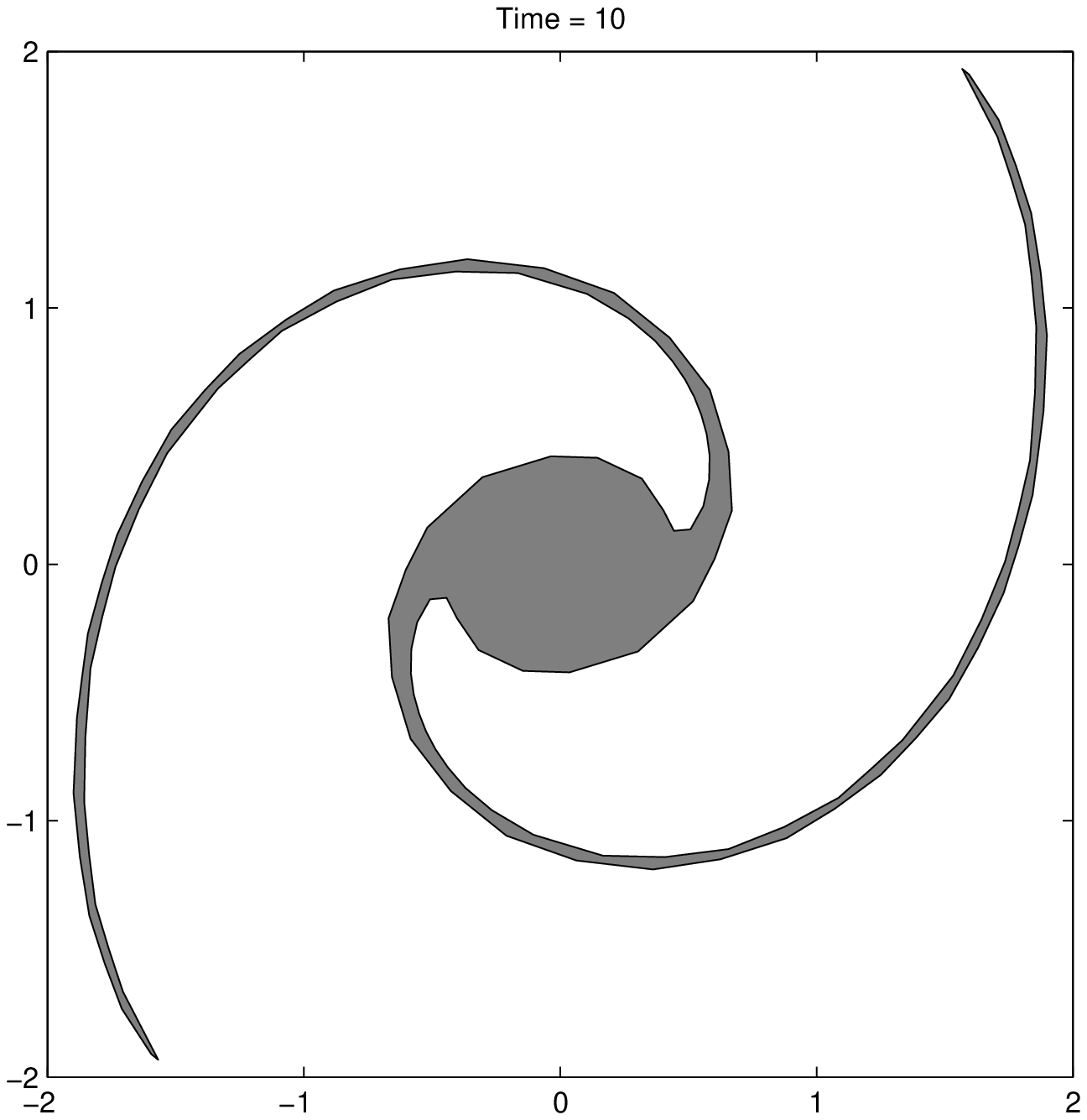}}}
\caption{Evolution of a swarm patch under the model (\ref{eq:2dincomp}). The  scalar interaction function $N$ is the Gaussian given by (\ref{eq:Gaussian}) with interaction length scale $d=1$. The constant population density is $\rho_0 = 1$. The initial shape is an ellipse with a major semiaxis of $1$ and a minor semiaxis of $0.1$. The patch eventually evolves into a circular core with two spiral arms.}
\label{fig:ellipseseries}
\end{figure}

The evolution is qualitatively similar even for swarm patches whose initial shape is far from circular. For instance, Figure \ref{fig:ellipseseries} shows the evolution of a swarm patch, again with the interaction function (\ref{eq:Gaussian}) and $d=1$. The boundary of the initial shape is an ellipse with a major semiaxis of $1$ and a minor semiaxis of $0.1$. As with the previous example, the swarm patch rotates in a clockwise direction. Spiral arms develop at the points furthest from the center of the patch, and there is a movement of population density towards a developing circular core, which is noticeable at time $t=8$ and well-defined by $t=10$.

\begin{figure}
\centerline{\resizebox{\textwidth}{!}{\includegraphics{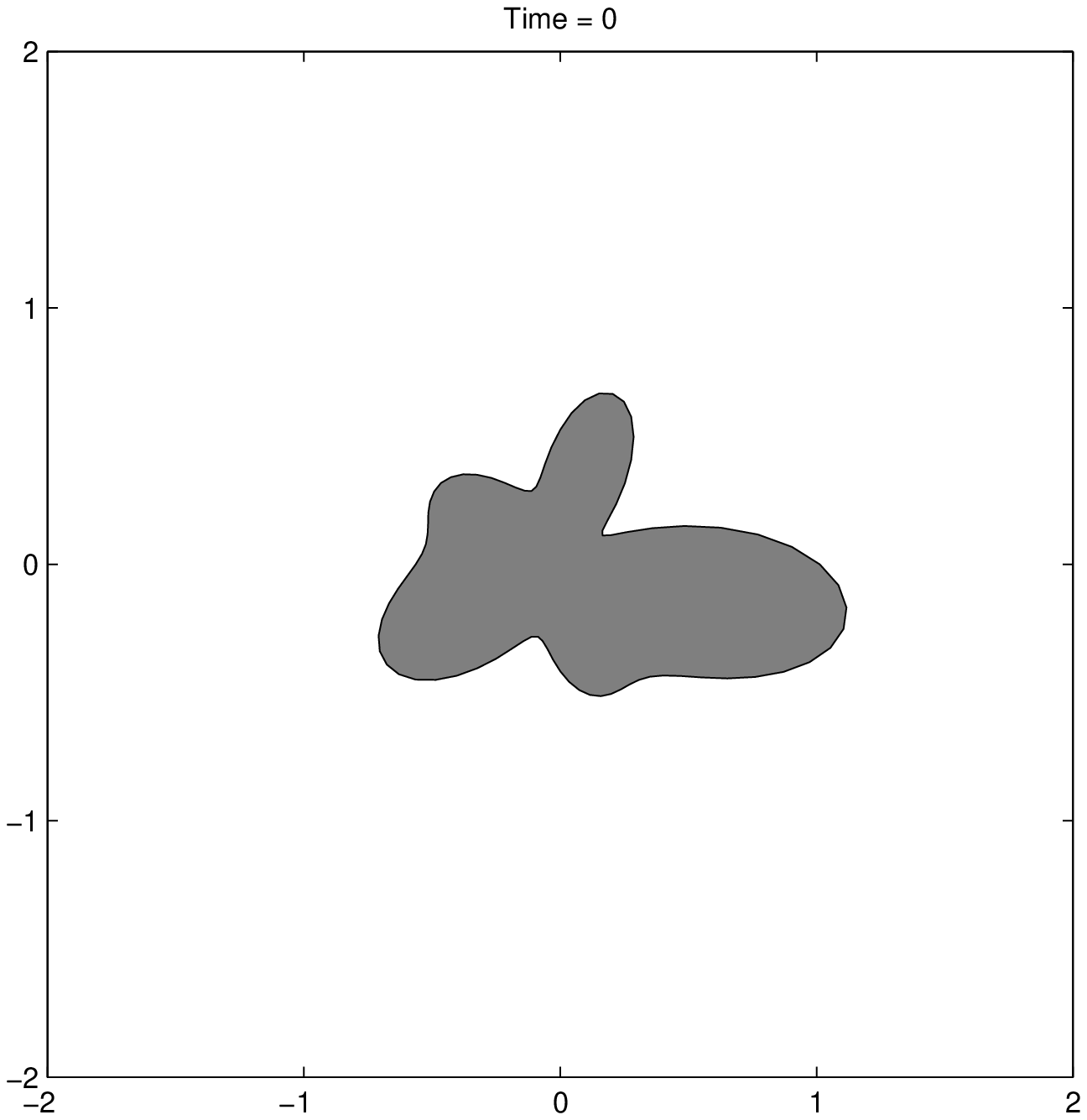} \hfill \includegraphics{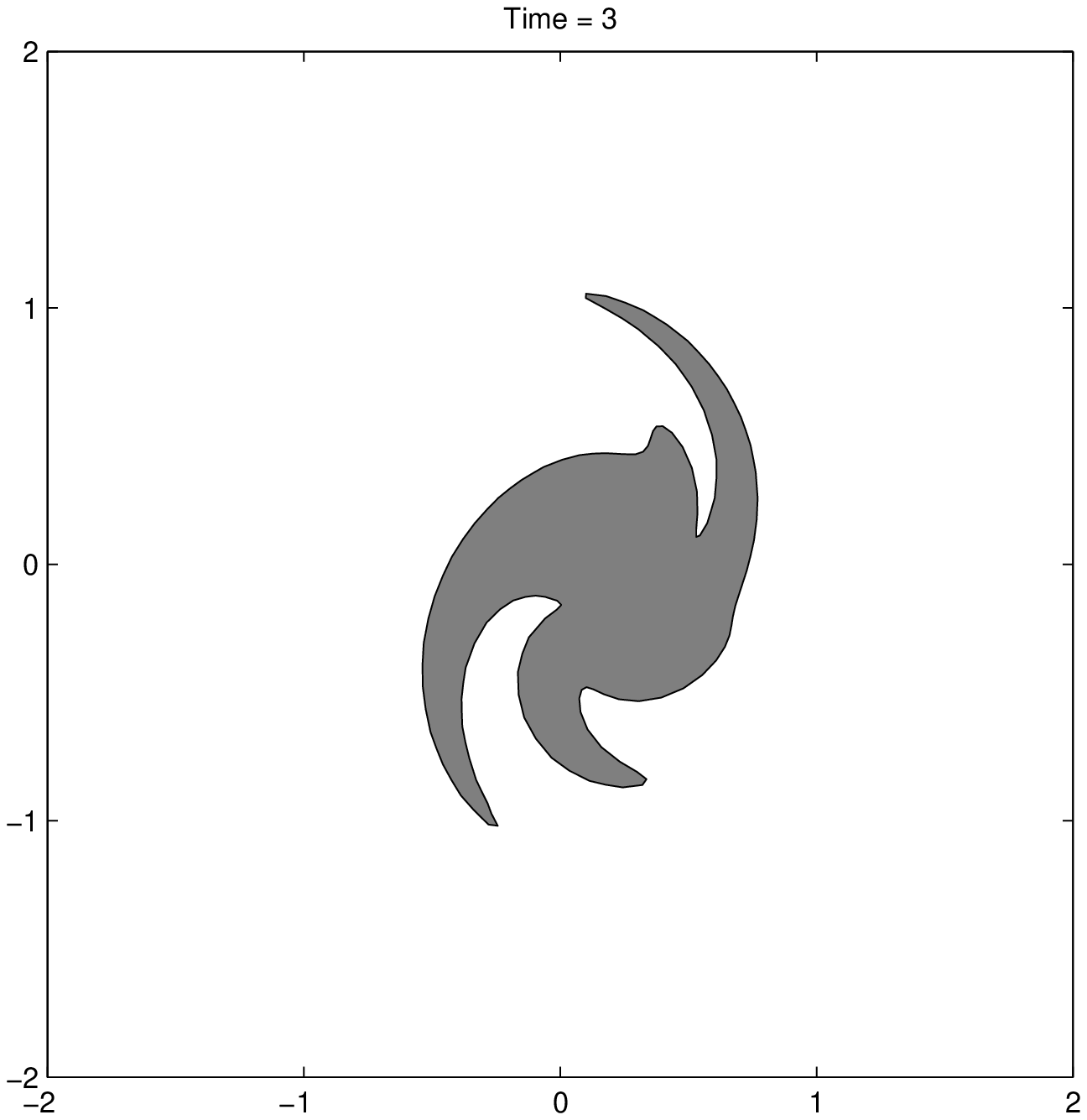}}}
\centerline{\resizebox{\textwidth}{!}{\includegraphics{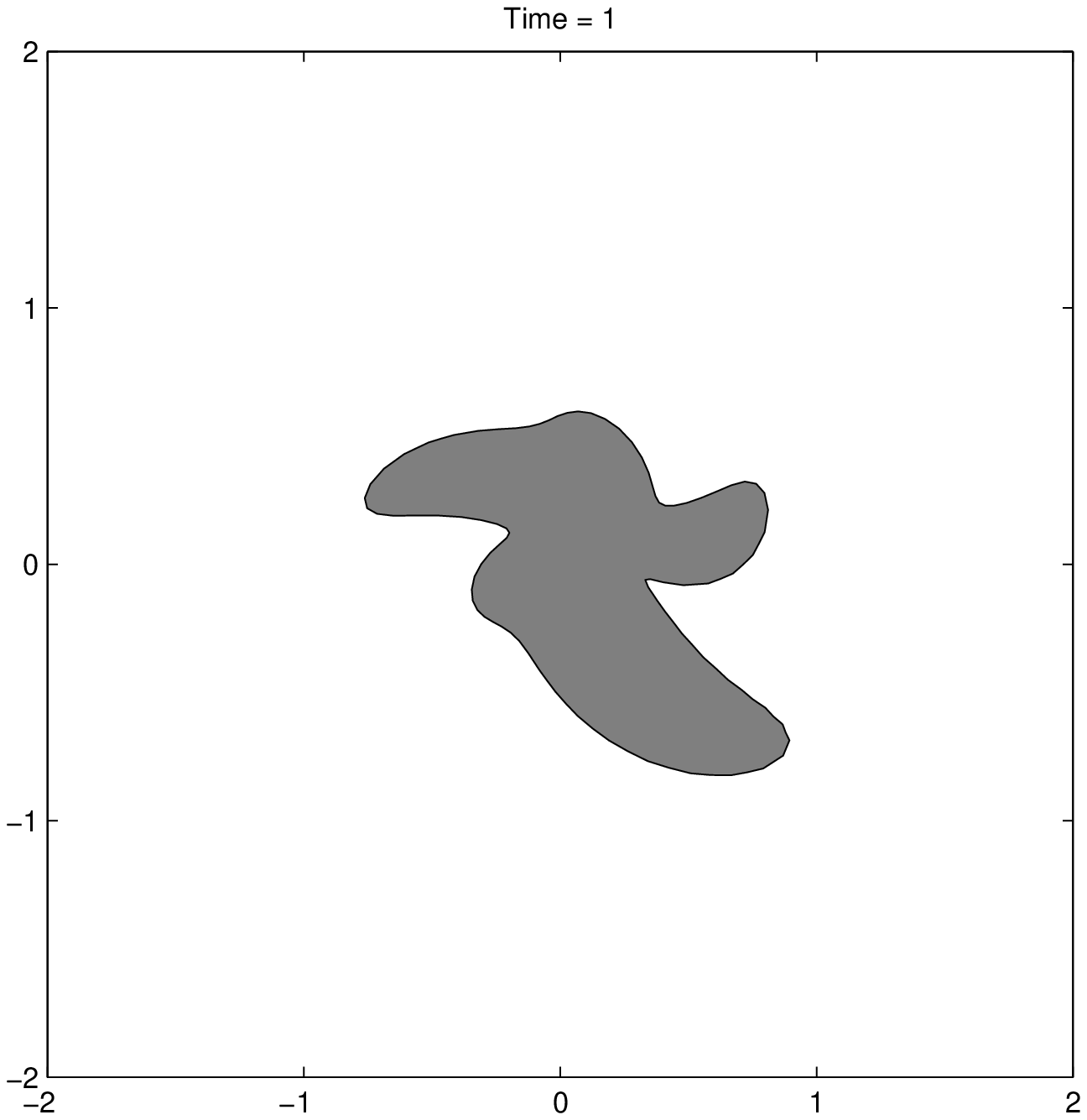} \hfill \includegraphics{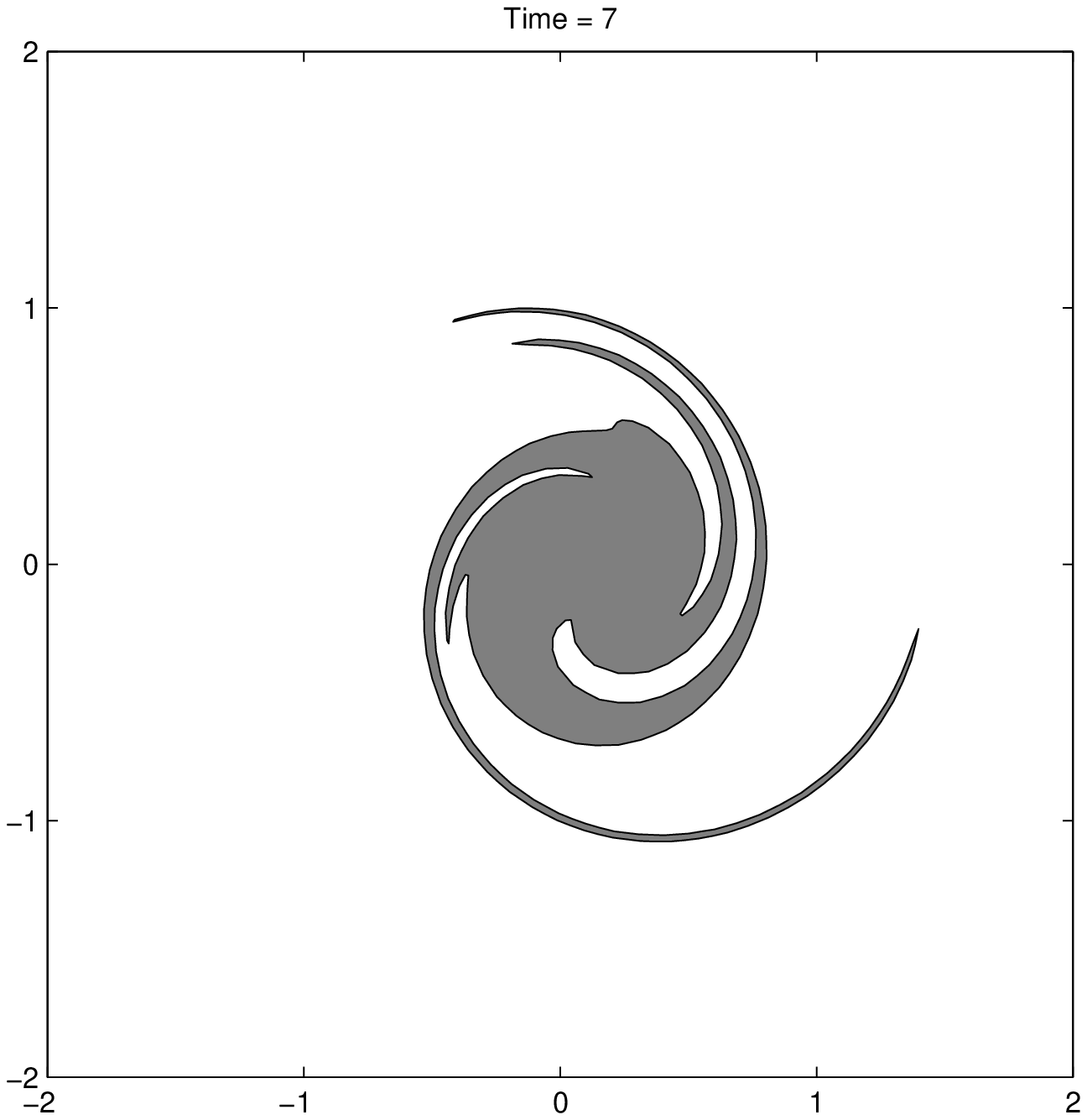}}}
\centerline{\resizebox{\textwidth}{!}{\includegraphics{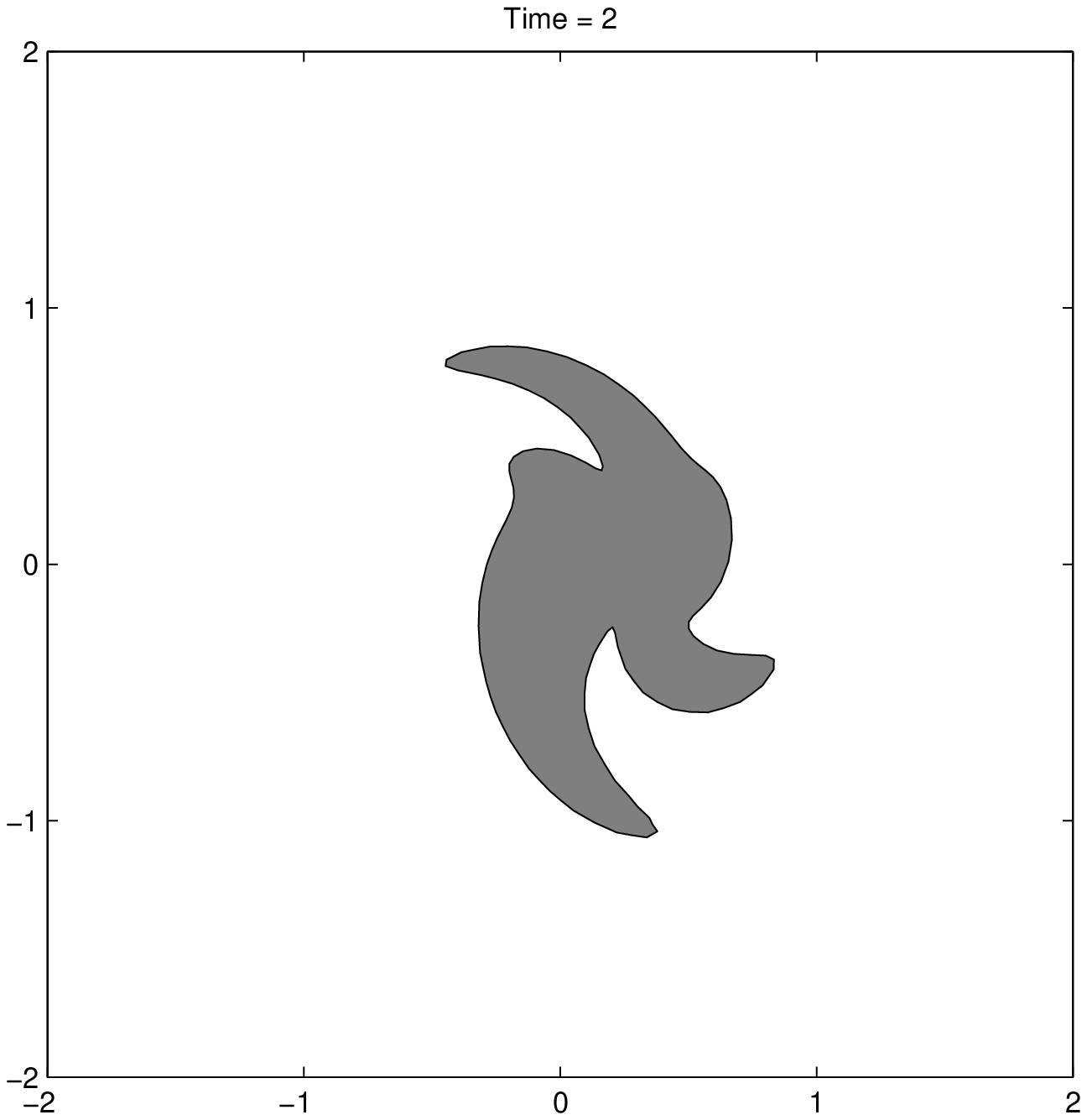} \hfill \includegraphics{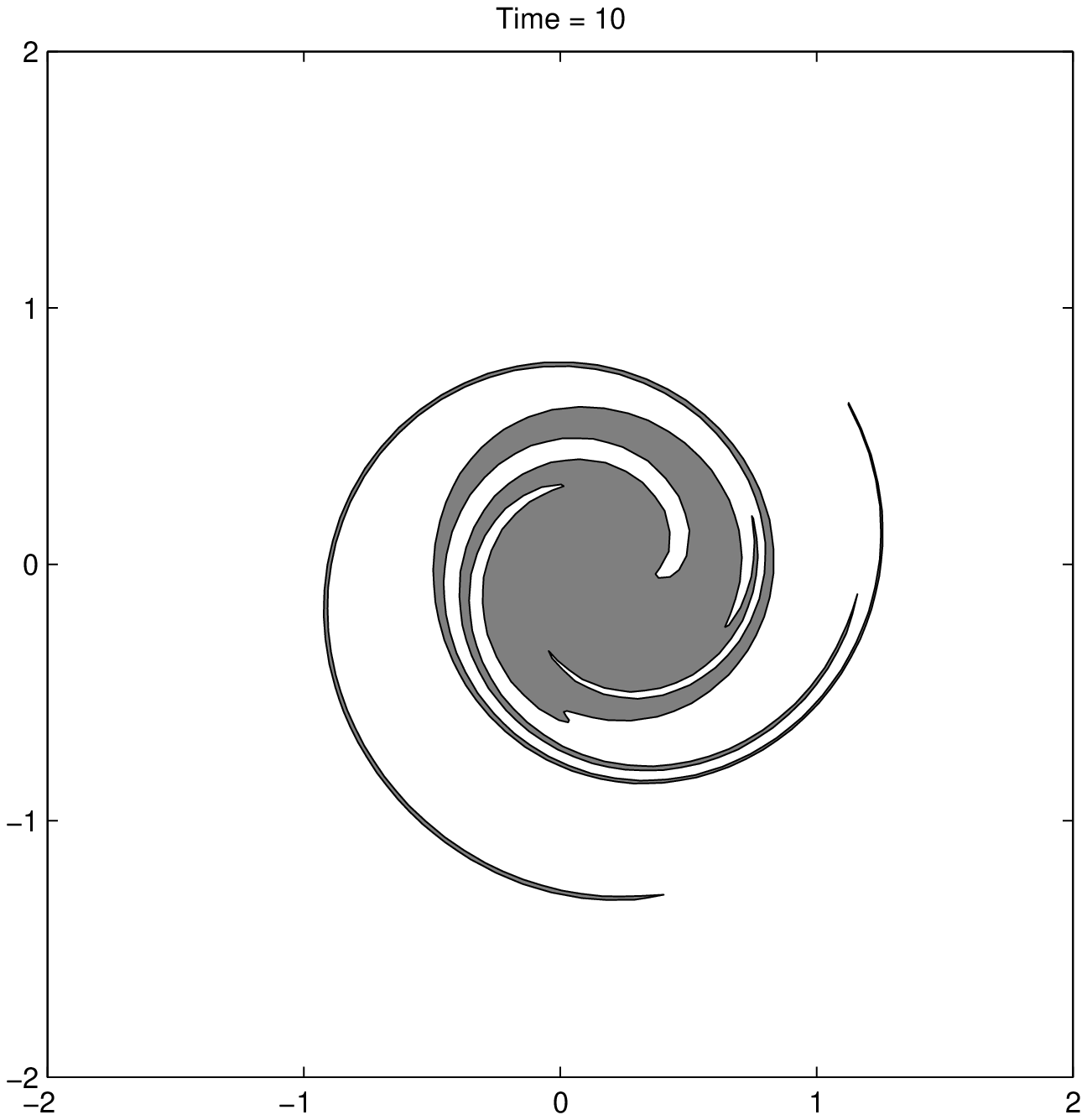}}} \caption{Evolution of a swarm
patch under the model (\ref{eq:2dincomp}). The scalar interaction function $N$ is the Gaussian given by (\ref{eq:Gaussian}) with interaction length scale
$d=1$. The constant population density is $\rho_0 = 1$. The initial shape is generated by the polar function $r = f(\theta)$ where $f$ is a
superposition of cosine components with randomly chosen amplitudes and randomly chosen low-integer frequencies. The patch eventually evolves
into a circular core with an irregular arrangement of spiral arms. We begin the simulation with 84 Lagrangian nodes on the swarm patch boundary 
and end with 485 nodes.} \label{fig:randomseries}
\end{figure}

Finally, we comment that a similar evolution occurs even for irregularly shaped swarm patches. Figure \ref{fig:randomseries} shows the evolution
of such a patch with the same interaction function as in the previous two examples. The initial shape is generated by the polar function $r =
f(\theta)$ where $f$ is a superposition of cosine components with randomly chosen amplitudes and randomly chosen low-integer frequencies. As
with the previous examples, the patch rotates clockwise, developing a circular core and an irregular arrangement of spiral arms.

We have shown in this section that in the case of incompressible motion, our simple nonlocal kinematic model has constant density solutions of compact support. It was seen directly from (\ref{eq:2dincomp}) that the evolution of any initial swarm patch slows for very large or very small values of the interaction length scale $d$. For intermediate values of $d$, numerical simulations demonstrated that the evolution of these swarm patches is rotational, with the direction of motion set \emph{a priori} by the sign on the interaction function~$N$. There is a flux of population towards the rotational center of the swarm, where a circular core develops. Spiral arms form at regions of the boundary where the curvature is very high. We saw that all of our numerical simulations resulted in asymptotic vortex states.

We close this section by discussing the biological significance of incompressible velocities. For a constant density swarm in which attractive and repulsive effects are in balance, organisms wishing to maintain constant density \emph{must} move with incompressible velocities. Potential velocities, discussed in the next section, will lead to variations in population density. Thus, one may think of the incompressible velocity terms as those which model the aggregate, cooperative dynamics of organisms striving to maintain equal spacing. Of course, equal spacing might also be maintained by means of a constant local drift, but this is not a cooperative effect, and will not lead to the type of vortex-like structure seen here.

\section{Potential motion}
\label{sec:potential}

In this case,
\begin{equation}
\vec{K}=\nabla P \label{eq:potkernel} 
\end{equation}
so that the model (\ref{eq:2dmodel1}) - (\ref{eq:2dmodel2}) may be written compactly as
\begin{equation}
\rho_t + \nabla \cdot [\rho(\nabla P * \rho)] = 0. \label{eq:2dpotential}
\end{equation}
We take the scalar interaction function $P=\mp G_d$ where $G_d$ is the Gaussian distribution of width $d$ given by (\ref{eq:Gaussian}). In Section 
\ref{sec:incomp}, we made the same choice for the scalar interaction function $N$. In that case, the sign of $N$ simply determined the direction of 
rotation of the swarm. For the present case, we will see that the sign of $P$ has a much more dramatic effect on the evolution of the population. 
Specifically, it will determine whether organisms disperse or aggregate, as we discuss in the following two subsections.

\subsection{Dispersion}
Here, we take $P = -G_d$. Note that (\ref{eq:2dpotential}) has an analogy to Darcy's law for flow in porous media. In fact, in the limit $d
\rightarrow 0$, $-G_d$ becomes a $\delta$-function of strength $\pi$ and the governing equation (\ref{eq:2dpotential}) is a the porous media
equation. This is a well-studied PDE which possesses an exact self-similar solution, called Barenblatt's solution. For an initial population of size $Q$ placed at the origin, Barenblatt's solution is
\begin{equation}
\rho(r,t) = \begin{cases}
\frac{1}{2\pi}\sqrt{\frac{Q}{t}} - \frac{r^2}{8\pi t} &\text{$r  \leq 2 \left( Qt\right)^{1/4} $}\\
0 &\text{$r  > 2 \left( Qt\right)^{1/4} $.}
\end{cases}
\label{eq:barenblatt}
\end{equation}
A discussion of the porous media equation as it relates to biological dispersal, along with a more general statement of Barenblatt's solution, may be found in~\cite{m2002}. In the opposite limit $d
\rightarrow \infty$, \emph{i.e.} when social interactions are extremely nonlocal, a bit of algebra again reveals that the equation becomes simply the
steady state $\rho_t=0$. The intuitive statement of this limiting case is similar to that in the previous section: organisms can sense
population gradients infinitely far away, but these gradients have no influence on velocity because the strength of the social interactions is infinitely weak.

For intermediate values of the interaction length scale $d$, the population density profile experiences diffusion and convection. We may understand this better by writing the governing equation (\ref{eq:2dpotential}) in an alternate form. After some algebra, we obtain
\begin{equation}
\rho_t = \nabla \rho \cdot (\nabla G_d * \rho) + \rho(\nabla^2 G_d * \rho) \label{eq:altpot}
\end{equation}
The first term on the right-hand side of (\ref{eq:altpot}) is convective, and due to the single derivative on $G$, scales like $1/d^4$. In contrast,
the second term is diffusive, and scales like $1/d^6$. Thus, for a given population density profile, as $d$ is increased, we expect that convection will be more dominant than diffusion.

We demonstrate the role that the interaction length scale $d$ plays by means of numerical simulations. For these examples, we focus on a radially
symmetric model, so that the density $\rho(r,t)$ is a function of the radial coordinate $r$ and the velocity $v$ is as given above.  Note that if
$\rho$ is radially symmetric and $P$ is also radially symmetric then the velocity field for this gradient flow points in the radial direction and is
itself radially symmetric. We solve the governing equation on the unit disk with boundary condition $\rho=0$ on the circumference. We use
MacCormack's method, which is second-order accurate in space and time; see, for instance, \cite{s1989}. We use $n=64$ grid
points with timesteps of $\Delta t = 1 \times 10^{-5} - 5 \times 10^{-4}$. Checks are performed with finer meshes in space and time to verify that solutions are sufficiently
well-converged.

\begin{figure}
\centerline{\resizebox{\textwidth}{!}{\includegraphics{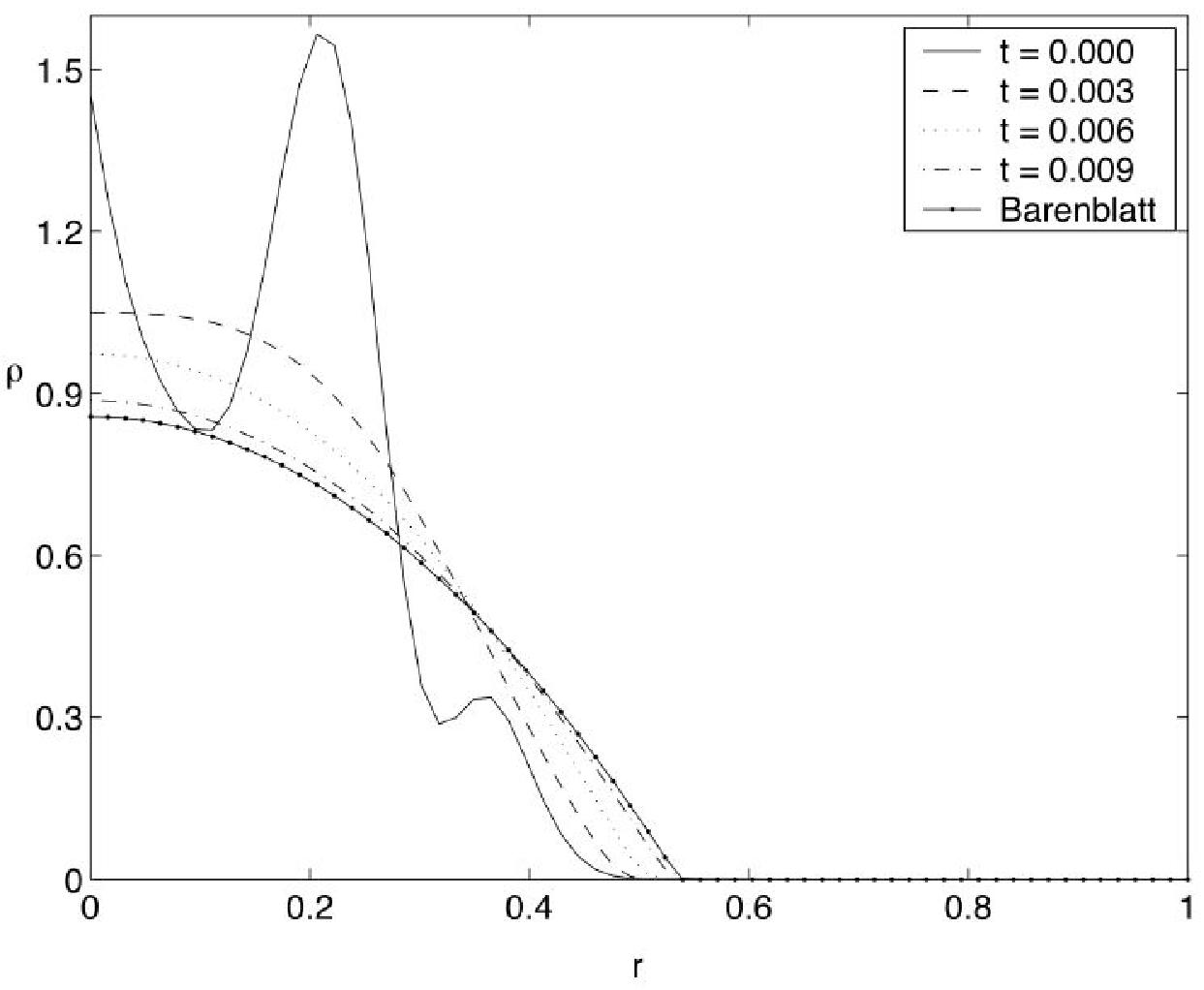}}}
\centerline{\resizebox{\textwidth}{!}{\includegraphics{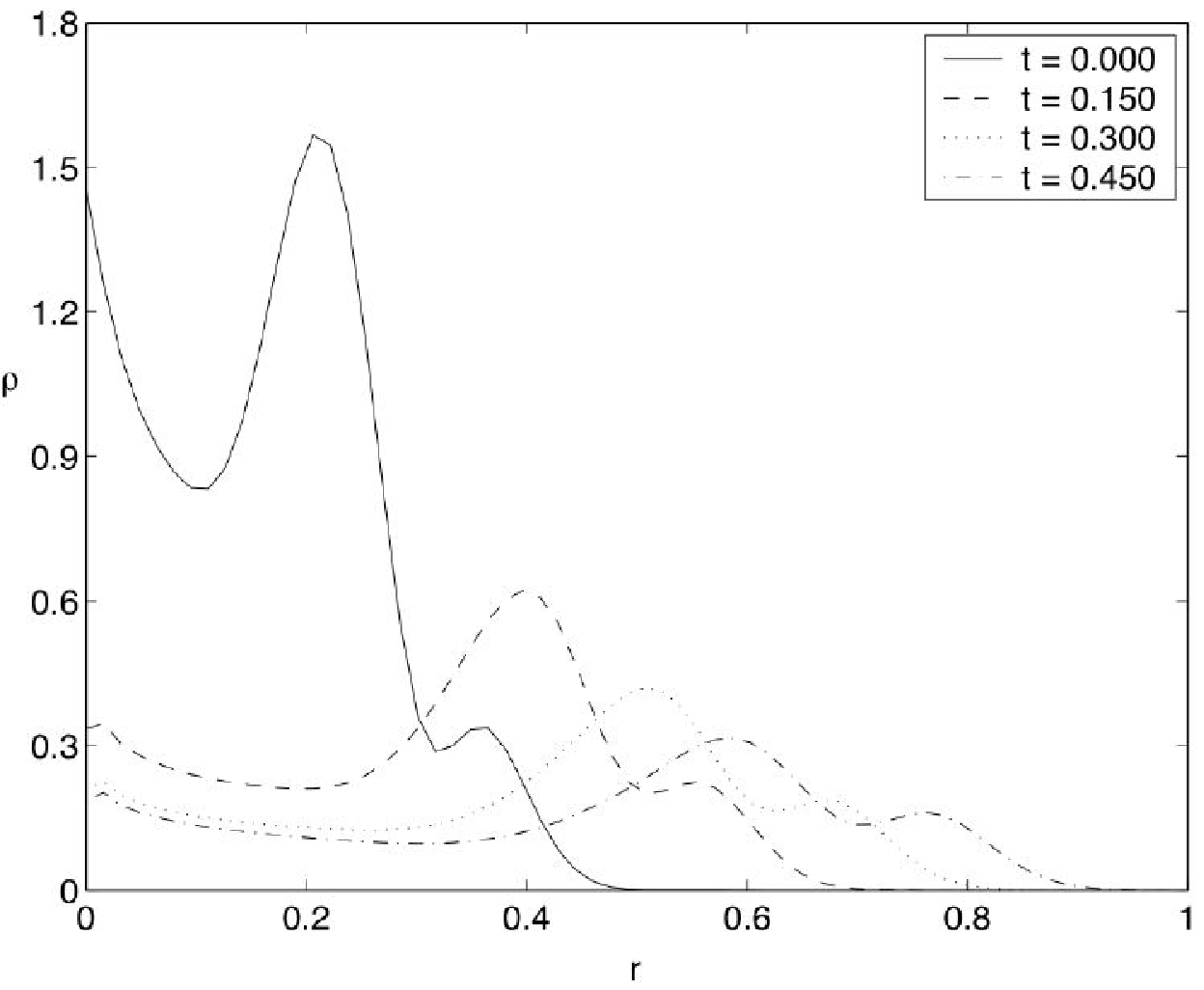}}}
\caption{Four snapshots of the time evolution of (\ref{eq:2dpotential}) in radially symmetric geometry with a random initial condition and the potential interaction function $P=-G_d$ given by (\ref{eq:Gaussian}). The graphs show the population density $\rho$ versus the radial coordinate $r$. with (a) $d = 0.01$ Interactions are very localized, and the dynamics are similar to those of the porous media equation. The curve labeled ``Barenblatt'' is a snapshot of the self-similar solution to the porous media equation given by (\ref{eq:barenblatt}) and should be compared to the numerical solution at time $t=0.009$. See text for details. (b) $d = 0.5$. For this case of more nonlocal interactions, the population is convected away from the origin and the smoothing of the population density profile occurs more slowly.}
\label{fig:dispersion}
\end{figure}

Two example simulations are shown in Figure \ref{fig:dispersion}. We choose as a random initial condition the function
\begin{equation}
\rho(r,0) = f(r)\left\{  \frac{1}{2}+\frac{1}{2}\tanh(5-15r) \right\}
\end{equation}
Here, $f(r)$ is created by superposing Fourier modes with low integer wave numbers and random coefficients. Multiplication by the bracketed combination is carried out so that the initial condition decays smoothly towards zero. For the first example, this state is evolved with $d=0.01$, so that social interactions are only 
very slightly nonlocal. In this case, (\ref{eq:2dpotential}) is nearly the porous media equation, and the ``bumpy'' initial condition quickly smooths out and approaches the parabolic profile given by Barenblatt's solution. We verify that the numerical solution approaches Barenblatt's solution as follows. We fit the numerical solution at time $t=0.006$ to  (\ref{eq:barenblatt}). The numerical solution is evolved numerically, and the fit Barenblatt's solution is evolved analytically. The two are compared again at time $t=0.009$. Both curves are contained in Figure \ref{fig:dispersion}a. The curves nearly overlay each other, and the maximum error between the two is $3\%$.

For contrast, we have taken the same random initial condition and integrated it with the more nonlocal interaction length scale $d=0.5$. In this case, Fourier modes are damped much more slowly, and the bumpy initial conditions retains its shape much longer. The motion of the swarm is much more convective, and the population is transported away from the origin.

\subsection{Aggregation}
\label{sec:aggregation}

In this case, we take $P = G_d$. Whereas $P$ was strictly negative in the previous subsection, it is now strictly positive, and this change has dramatic
consequences for the dynamics. Now, the governing equation states that velocities are up, rather than down, population gradients
(nonlocally), so that the population will tend to form groups.

We may understand this grouping by means of a linear stability analysis. To do so, we consider small perturbations $\hat{\rho}$ to a constant density steady state $\rho_0$. Linearizing (\ref{eq:2dpotential}), we obtain
\begin{equation}
\hat{\rho}_t =  -\rho_0 G_d * \nabla^2 \hat{\rho} 
\label{eq:nonlocalheat}
\end{equation}
and thus we see that the perturbation obeys a nonlocal backwards heat equation. Taking a Fourier \emph{ansatz} for the perturbation \emph{i.e.} $\hat{\rho} = \hat{\rho}_0 e^{i(\vec{k} \cdot \vec{x} + \sigma t)}$, we find that the linear growth rate is given by
\begin{equation}
\sigma(k) = \pi \rho_0 k^2 e^{-k^2/(4d^2)}.
\label{eq:growthrates}
\end{equation}
where $k=|\vec{k}|$.
By computing the critical points of (\ref{eq:growthrates}) we see that the most unstable modes are those with wave number $k_u=2/d$. The growth of this most
unstable mode provides a mechanism for the clumping of organisms. We expect that extremely localized interactions will lead to the formation of 
a larger number of small groups, \emph{i.e.} a density distribution pattern with a small characteristic length scale. On the other hand, more nonlocal interactions will result in a smaller number of large groups, \emph{i.e.} a density distribution pattern with a larger characteristic length scale.

We confirm this prediction by means of numerical simulations. Using a pseudospectral Fourier method with 128 modes on each axis, we integrate (\ref{eq:2dpotential}) on a $2\pi \times 2\pi$ box with periodic boundary conditions.  We choose the initial density distribution to be $\rho = 1$ plus a small random perturbation constructed by superposing low wavenumber ($k<15$) Fourier modes with random coefficients. For time-stepping, we initialize with a forward Euler step and then use a second-order Adams-Bashforth method. Depending on the value of the interaction length scale $d$, we take time steps of $\Delta t = 4 \times 10^{-5} - 1 \times 10^{-3}$. Checks are performed with different numbers of modes and different time steps to verify convergence.

\begin{figure}
\centerline{\resizebox{\textwidth}{!}{\includegraphics{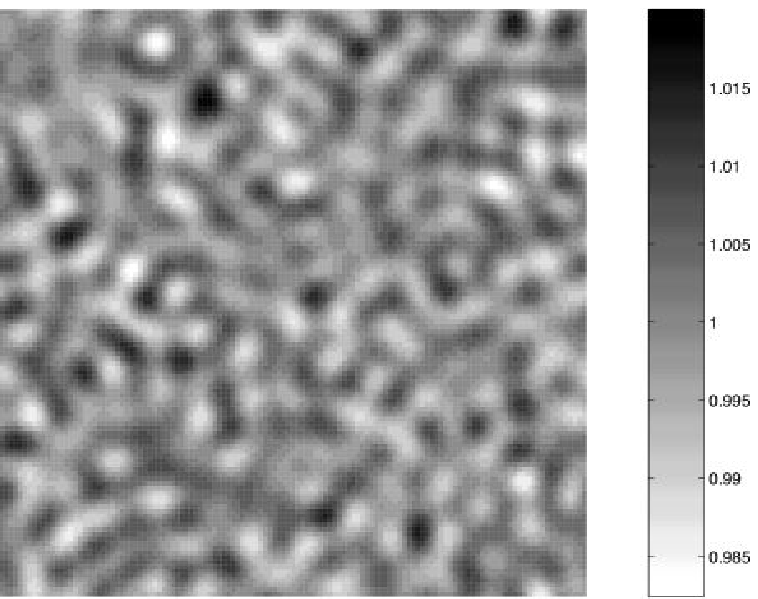} \hspace{0.5in} \includegraphics{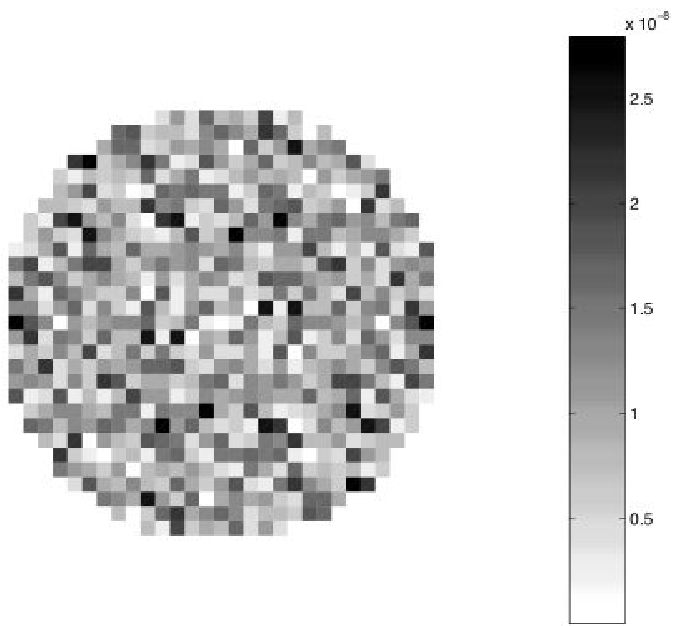}}}
\centerline{\resizebox{\textwidth}{!}{\includegraphics{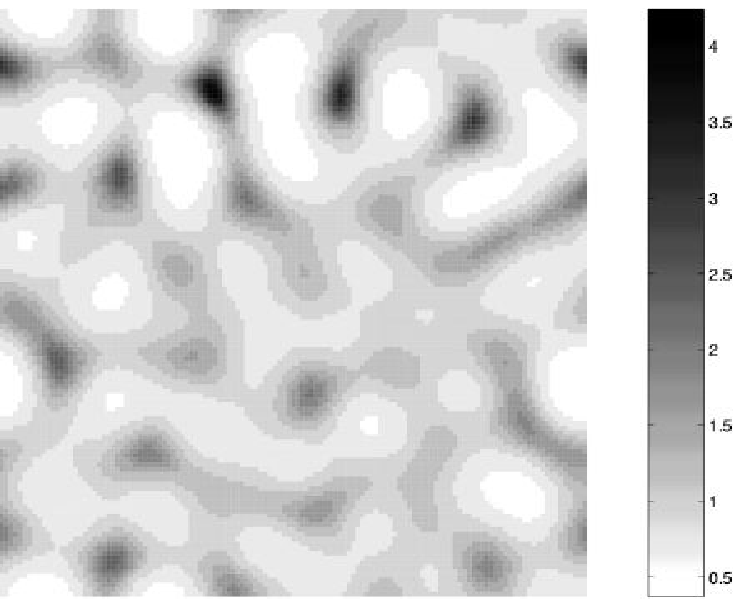} \hspace{0.5in} \includegraphics{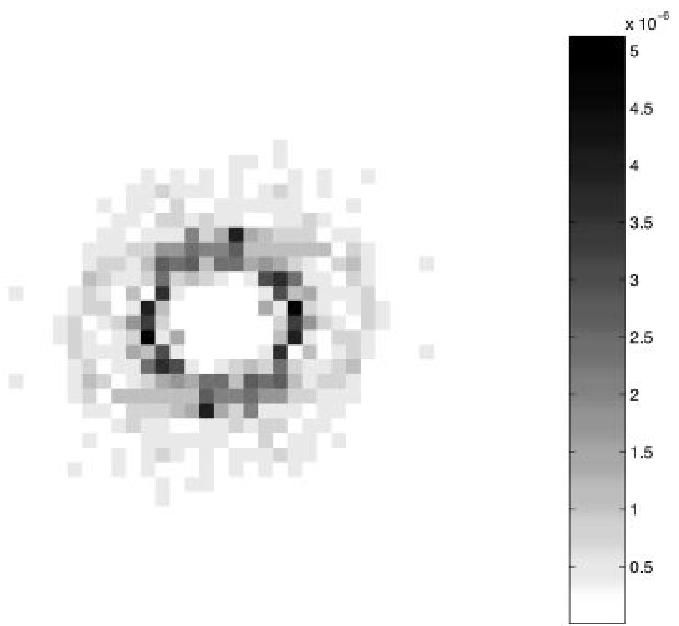}}}
\centerline{\resizebox{\textwidth}{!}{\includegraphics{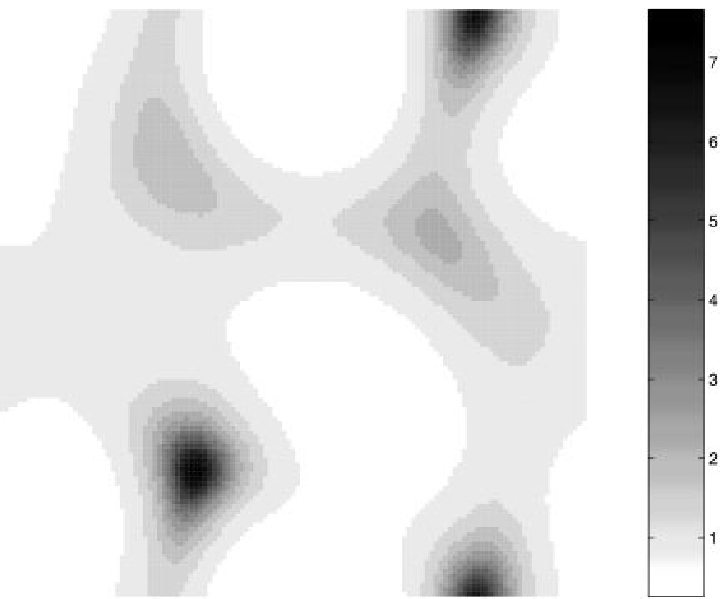} \hspace{0.5in} \includegraphics{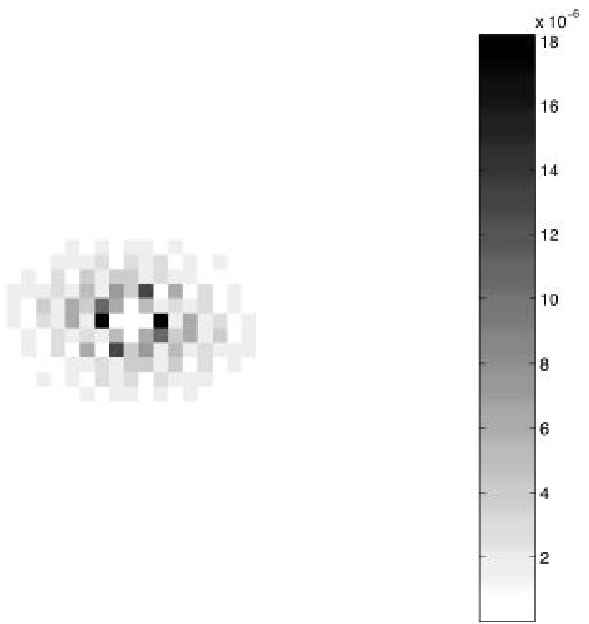}}}
\caption{Results from integrating (\ref{eq:2dpotential}) on a $2\pi \times 2\pi$ periodic box with interaction function $P=G_d$ given by 
(\ref{eq:Gaussian}). (a) Initial population density, given by $\rho=1$ plus a small random perturbation. (b) Center of the power spectrum of the initial 
perturbation. (c)~Population density at $t=0.132$ with interaction length scale $d=0.4$. Note the formation of small, high-density groups. (d) 
Center of the power spectrum of the perturbation at $t=0.132$. The strongest peaks are at $k=5$, which is the most unstable mode as predicted by linear 
analysis. Figures (e) and (f) are analogous to (c) and (d), but data is taken at $t=2.74$ and the longer interaction length scale $d=1$ is 
used. In this more nonlocal case, the most unstable wave number is $k=2$ and the population clumps into fewer, larger groups.}
\label{fig:aggregation}
\end{figure}

Our results are shown in Figure \ref{fig:aggregation}. Figures \ref{fig:aggregation}a shows the initial condition. Dark patches correspond to regions of
higher density. Figure \ref{fig:aggregation}b shows the center of the power spectrum of the initial perturbation, which is noisy. Figure \ref{fig:aggregation}c
shows the evolution of the state in Figure \ref{fig:aggregation}a at time $t=0.132$ with interaction length scale $d=0.4$.  Notice the patches of high
population density. By the linear stability arguments given above, the characteristic wave number of the grouping pattern is predicted to be
$k_u=2/d=5$. Figure \ref{fig:aggregation}d shows a blow-up of the center part of the power spectrum of the evolution of the perturbation. As 
predicted,
the strongest peaks are centered around the circle $k=5$. Figures \ref{fig:aggregation}e and \ref{fig:aggregation}f are analogous to \ref{fig:aggregation}c and
\ref{fig:aggregation}d, but at time $t=2.74$ and with the more nonlocal interaction length scale $d=1$. In this case, fewer groups form, and they are
larger. The most unstable wave number from linear analysis is $k_u=2$, and indeed this is the wave number corresponding to the strongest peaks in the
power spectrum. Finally, we comment that these simulations are not continued for longer times because they experience exponential blow up, due to
the lack of any effects to counterbalance the attractive forces in the model. See the appendix for a mathematical discussion of the 
blow-up. Modifications to the model to prevent blow up will be a key aspect of future work, as mentioned in the next section.

\section{Conclusions}
\label{sec:conclusions}

Our work on biological groups in two dimensions is in 
the spirit of the one dimensional study in \cite{mk1999}. The 
overarching goal in this paper is to make specific statements 
about how social interactions between organisms affect the large-scale 
motion of a biological group.  In summary, we formulated and studied a 
simple 
kinematic continuum model which includes nonlocal, spatially-decaying 
social interactions between individuals. We decomposed the 
dynamics of our model into incompressible motion and potential motion, 
or alternatively, motion perpendicular to population gradients and 
motion along gradients.

Whereas a constant drift and cancellation of nonlocal effects are 
necessary for maintaining a cohesive swarm in one-dimension, this is 
not the case in two dimensions. For the special case of an 
incompressible kernel, so that organisms move perpendicular to 
population gradients (in a nonlocal sense), the equations have constant
density solutions of compact support.  Through numerical simulations 
beginning with a variety of initial conditions, we showed that the 
dynamics for incompressible interactions are rotational, and that swarm 
patches eventually develop vortex-like structure. This rotational motion 
is a cooperative mechanism by which  a swarm may maintain motion and 
cohesion once potential (attractive and repulsive) effects have come 
into balance. The sign of the social interaction term determines the 
direction of rotation, and the social
interaction length scale determines the degree of macroscopic group 
movement. The observed asymptotic vortex states are intriguingly 
similar to actual  mill vortices seen in biological systems 
\cite{s1971,bccvg1997,pk1999,rnsl1999,obm2003}.

In contrast, potential kernels model repulsion or attraction between 
organisms. For the repulsive case, the interaction length scale 
determines a balance between diffusive motion and convective motion. 
Very localized interactions lead to greater smoothing, while more 
nonlocal interactions result in slower smoothing, but more outward
convection motion. For the attractive case, the length scale determines 
a most-unstable mode, the growth of which results in the clumping of 
the population into regions of high and low density.

This work leaves open many possibilities for future research. One route 
would be to conduct fully two-dimensional simulations of biological 
groups under the simultaneous influence of incompressible and potential 
interactions. Another would be to relax some of the simplifying 
modeling assumptions made in Section \ref{sec:2d}. The focus would be 
on more complicated velocity rules containing both nonlocal and local 
components, each of which may be nonlinear. Finally, future work might 
consider dynamic, rather than kinematic, velocity rules. These rules 
would take into account inertial forces
that might capture ``phase changes'' in animal group
behaviors such as the transition from milling to translational motion.
Ultimately, a model should base 
interaction rules on specific biological socialization functions of the 
organisms. Unfortunately, field and laboratory data leading to such 
models is very limited. We hope that the general discussion of this 
paper will help to focus further research in this direction.
\appendix

\section{Appendix}
\label{sec:appendix}

In this appendix we sketch proofs for two results mentioned in the body of this paper, namely regularity of a swarm-patch boundary 
for the model examined in Section \ref{sec:incomp} and an exponential upper bound for the blow-up of the model examined in Section \ref{sec:aggregation}.

\subsection{Regularity of the boundary of swarm patches}

In this subsection we discuss the swarm patch model of Section \ref{sec:incomp} and the
regularity of the swarm patch boundary.  The boundary parametrized by $\vec{z}(\alpha,t)$
is a solution of the integrodifferential equation
\begin{equation}
\frac{d\vec{z}}{dt} = \vec{v}(\vec{z},t), \quad \vec{v}(\vec{x},t) = \vec{K}*\rho(\vec{x},t), \quad \rho(\vec{x},t) = \chi_{\small{\substack{ \\ \Omega}}}\label{ODE}
\end{equation} where $\Omega(t)$ is the interior of the swarm patch.

If we consider the problem as an integrodifferential equation for $\rho$,
\begin{equation}
\frac{d\rho}{dt} + \vec{v}\cdot \nabla \rho = 0, \quad \vec{v} = \vec{K}*\rho
\end{equation}
where $\vec{K}=\nabla^\perp N$ for a smooth radial function
$N$ decaying at infinity, then the swarm patch is an example of a
weak solution of this problem with initial condition $\rho_0\in L^1\cap L^\infty (\mathbb{R}^2)$.
Existence and uniqueness of this problem can be proved following the
classical theory of Yudovich for vortex patches \cite{y1963} 
which is also detailed in \cite{mb2002}.  Such a discussion is
beyond the scope of this paper.  However, we present some straightforward
estimates that can be used to prove that the boundary of the patch
remains smooth if initially smooth, as in the case of vortex patches
for which the kernel $\vec{K}$ is more singular (and the proof is correspondingly
more difficult).

If the swarm density persists as the characteristic function
of a domain $\Omega (t)$ then it is uniformly bounded in $L^1\cap L^\infty$
for all time.  Since $\vec{v}=\vec{K}*\rho$ with $\vec{K}\in C^\infty$ then we have an \emph{a priori}
bound for all derivatives of $\vec{v}$, $D^k \vec{v}<\infty $ for all multi-indices $k$.
Smoothness of the patch boundary then follows from the fact that 
the map $\vec{z}$ satisfies the ODE (\ref{ODE}) with initial condition
\begin{equation}
\vec{z}|_{t_0}= \vec{z}_0(\alpha)
\end{equation}
for smooth $\vec{z}_0$.  Since $\vec{v}$ is $C^\infty$
by standard regularity theory for solutions of ODEs we see that $\vec{z}$ itself
is smooth.  Note that the mapping $\vec{z}$ can not develop a critical point
at a later time $t$ because the Lagrangian derivative of $\vec{z}$, 
$\vec{z}_\alpha$ satisfies the ODE 
\begin{equation}
\frac{d \vec{z}_\alpha}{dt} = \nabla \vec{v} \mid_{\vec{z}(\alpha,t)}  \vec{z}_{\alpha}.
\end{equation}
Since $\nabla \vec{v}$ is bounded for all time then $\vec{z}_\alpha$ remains
bounded away from zero and infinity if it is so bounded at time zero,
by Gr\"onwall's Lemma.

\subsection{Boundedness of the swarm density for a general velocity rule}

In Section \ref{sec:aggregation} we presented numerical computations that showed $\rho(\vec{x},t)$
can exhibit blowup when the convolution kernel $\vec{K}=\nabla P = \nabla G_d$ is positive.
In the limit as $d \rightarrow 0$ this formally corresponds to a backward
time porous media equation.

Here we derive an \emph{a priori} bound that shows that for a smooth kernel
of any sign, that the maximum of $\rho$ is bounded by an exponential in time.
Thus the blowup seen numerically must be an infinite time blowup, not
finite time.  The bound we derive depends on the $L^\infty$ norm
of the convolution kernel.  Thus as $d \rightarrow 0$ the bound 
itself becomes unbounded, as it should because we are approaching
the ill-posed limit in the positive kernel case.

In Eulerian coordinates, $\rho$ satisfies a reaction convection
equation
\begin{equation}
\rho_t + \vec{v} \cdot \nabla \rho = -\rho \nabla \cdot \vec{v}.
\end{equation}
This problem can be transformed to Lagrangian coordinates
using the method of characteristics.  Let $\vec{X}(\alpha,t)$ denote
the solution of the ODE
\begin{equation}
\frac{d\vec{X}}{dt} = \vec{v}(\vec{X},t), \quad \vec{X}|_{t=0} = \alpha.
\end{equation}
Then in the Lagrangian coordinate $\alpha$, 
$\rho$ satisfies
\begin{equation}
\frac{d \rho}{d t} = -(\nabla \cdot \vec v)|_{\vec{X}(\alpha, t)} \rho.
\end{equation}
Thus we have a differential inequality for the $L^\infty$ norm of $\rho$,
\begin{equation}
\frac{d}{dt}\|\rho\|_{L^\infty} \leq C \| \nabla \cdot \vec{v}\|_{L^\infty} \| \rho\|_{L^\infty}.\label{eq:dtrho}
\end{equation}
Since $\rho$ is a density, it has an \emph{a priori} $L^1$ bound,
\begin{equation}
\int \rho(\vec{x},t) d\vec{x} = \int \rho (\vec{x}, 0) d\vec{x} = \|\rho\|_{L^1}.
\end{equation}
Since $\vec{v}=\vec{K}*\rho$ for smooth $\vec{K}$, we have
\begin{equation}
\|\nabla \cdot \vec{v}\|_{L^\infty}\leq \|\nabla \cdot \vec{K}\|_{L^\infty} \|\rho\|_{L^1}.
\end{equation}
Combining this with (\ref{eq:dtrho}) and the a priori bound on the $L^1$
norm of $\rho$, we have
\begin{equation}
\frac{d}{dt}\|\rho\|_{L^\infty} \leq C \|\rho\|_{L^1} \| \rho\|_{L^\infty}, \quad C = \|\nabla \cdot \vec{K}\|_{L^\infty}.
\end{equation}
Gr\"onwall's Lemma then gives 
\begin{equation}
\|\rho\|_{L^\infty} \leq exp (C \|\rho_0\|_{L^1}t) \|\rho_0\|_{L^\infty}.
\end{equation}

\begin{acknowledgement}
This research is supported by NSF grants DMS-9983320 and DMS-0074049, ARO grant DAAD-19-02-1-0055, and ONR grant N000140310073. We are especially indebted to Daniel Marthaler for his input. We also acknowledge helpful discussions with Anke Ordemann.
\end{acknowledgement}

\bibliographystyle{elsart-num}
\bibliography{bibliography}

\begin{thebibliography}{10}
\expandafter\ifx\csname url\endcsname\relax
  \def\url#1{\texttt{#1}}\fi
\expandafter\ifx\csname urlprefix\endcsname\relax\def\urlprefix{URL }\fi

\bibitem{ah1990}
W.~Alt, G.~Hoffman (Eds.), Biological Motion: Proceedings of a Workshop held in
  K{\"{o}}nigswalter, Germany, March 16 - 19, 1989, no.~89 in Lecture Notes in
  Biomathematics, Springer-Verlag, Berlin, 1990.

\bibitem{ph1997}
\oneletter{J.K.} Parrish, \oneletter{W.M. Hamner} (Eds.), Animal Groups in
  Three Dimensions, Cambridge University Press, Cambridge, UK, 1997.

\bibitem{ogk1999}
A.~Okubo, D.~Grunbaum, L.~Edelstein-Keshet, The dynamics of animal grouping,
  in: A.~Okubo, S.~Levin (Eds.), Diffusion and Ecological Problems, 2nd
  Edition, Vol.~14 of Interdisciplinary Applied Mathematics: Mathematical
  Biology, Springer, New York, 1999, Ch.~7, pp. 197--237.

\bibitem{k2001}
L.~Edelstein-Keshet, Mathematical models of swarming and social aggregation,
  in: Proceedings of the 2001 International Symposium on Nonlinear Theory and
  its Applications, 2001, pp. 1--7.

\bibitem{vcbcs1995}
T.~Vicsek, A.~Czir{\'{o}}k, E.~Ben-Jacob, I.~Cohen, O.~Shochet, Novel type of
  phase transition in a system of self-driven particles, Phys. Rev. Lett.
  75~(6) (1995) 1226--1229.

\bibitem{tt1998}
J.~Toner, Y.~Tu, Flocks, herds, and schools: A quantitative theory of flocking,
  Phys. Rev. E 58~(4) (1998) 4828--4858.

\bibitem{lrc2001}
H.~Levine, \oneletter{W.J.} Rappel, I.~Cohen, Self-organization in systems of
  self-propelled particles, Phys. Rev. E 63 (2001) 017101.1--017101.4.

\bibitem{ckjrf2002}
I.~Couzin, J.~Krause, R.~James, \oneletter{G.D.} Ruxton, \oneletter{N.R.}
  Franks, Collective memory and spatial sorting in animal groups, J. Theor.
  Biol. 218~(1) (2002) 1--11.

\bibitem{ah2003}
M.~Aldana, C.~Huepe, Phase transitions in self-driven many-particle systems and
  related non-equilibrium models: A network approach, J. Stat. Phys. 112~(1--2)
  (2003) 135--153.

\bibitem{mkbs2003}
A.~Mogilner, L.~Edelstein-Keshet, L.~Bent, A.~Spiros, Mutual interactions,
  potentials, and individual distance in a social aggregation, in press (2003).

\bibitem{kwg1998}
L.~Edelstein-Keshet, J.~Watmough, D.~Grunbaum, Do travelling band solutions
  describe cohesive swarms? {A}n investigation for migratory locusts, J. Math.
  Bio. 36~(6) (1998) 515--549.

\bibitem{mk1999}
A.~Mogilner, L.~Edelstein-Keshet, A non-local model for a swarm, J. Math. Bio.
  38~(6) (1999) 534--570.

\bibitem{vcfh1999}
T.~Vicsek, A.~Czir{\'{o}}k, \oneletter{I.J.} Farkas, D.~Helbing, Application of
  statistical mechanics to collective motion in biology, Physica A 274~(1--2)
  (1999) 182--189.

\bibitem{s1971}
\oneletter{T.C.} Schnierla, Army Ants: A Study in Social Organization, W.H.
  Freeman, San Francisco, 1971.

\bibitem{bccvg1997}
E.~Ben-{J}acob, I.~Cohen, A.~Czir{\'o}k, T.~Vicsek, \oneletter{D.L.} Gutnick,
  Chemomodulation of cellular movement, collective formation of vortices by
  swarming bacteria, and colonial development, Physica A 238~(1--4) (1997)
  181--197.

\bibitem{pk1999}
\oneletter{J.K.} Parrish, L.~Edelstein-Keshet, Complexity, pattern, and
  evolutionary trade-offs in animal aggregation, Science 284 (1999) 99--101.

\bibitem{rnsl1999}
\oneletter{W.J.} Rappel, A.~Nicol, A.~Sarkissian, H.~Levine, Self-organized
  vortex state in two-dimensional {D}ictyostelium dynamics, Phys. Rev. Lett.
  83~(6) (1999) 1247--1250.

\bibitem{obm2003}
A.~Ordemann, G.~Balazsi, F.~Moss, Motions of {D}aphnia in a light field: Random
  walks with a zooplankton, Nova Acta Leopoldina To appear.

\bibitem{mb2002}
\oneletter{A.J.} Majda, \oneletter{A.L.} Bertozzi, Vorticity and Incompressible
  Flow, Texts in Applied Mathematics, Cambridge University Press, Cambridge,
  UK, 2002.

\bibitem{zhr1979}
\oneletter{N.J.} Zabusky, \oneletter{M.H.} Hughes, \oneletter{K.V.} Roberts,
  Contour dynamics for the {E}uler equations in two dimensions, J. Comput.
  Phys. 30~(1) (1979) 96--106.

\bibitem{bc1993}
\oneletter{A.L.} Bertozzi, P.~Constantin, Global regularity for vortex patches,
  Comm. Math. Phys. 152~(1) (1993) 19--28.

\bibitem{c1994}
\oneletter{J.Y.} Chemin, Persistency of geometric structures in bidimensional
  incompressible fluids, Ann. Sci. Ec. Norm. Sup. 26~(4) (1994) 517--542.

\bibitem{d1989}
\oneletter{D. G.} Dritschel, Contour dynamics and contour surgery: Numerical
  algorithms for extended, high-resolution modeling of vortex dynamics in
  two-dimensional, inviscid, incompressible fows, Computer Physics Reports 10
  (1989) 77--146.

\bibitem{m2002}
\oneletter{J.D.} Murray, Mathematical Biology I: An Introduction, 3rd Edition,
  no.~17 in Interdisciplinary Applied Mathematics, Springer, New York, 2002.

\bibitem{s1989}
\oneletter{J.C.} Strikwerda, Finite Difference Schemes and Partial Differential
  Equations, Chapman \& Hall, New York, 1989.

\bibitem{y1963}
V.~Yudovich, Non-stationary flows of an ideal incompressible fluid, Zh.
  Vychisl. Mat. Mat. Fiz. 3 (1963) 1032--1066.

\end{thebibliography}

\end{document}